\documentclass{aa}  

\usepackage{graphicx}
\usepackage{comment}
\usepackage{bm}
\usepackage{multirow}
\graphicspath{}
\usepackage{wrapfig}
\usepackage{acronym}

\usepackage{txfonts}
\usepackage{hyperref}
\hypersetup{
    colorlinks=true,
    linkcolor=blue,
    citecolor=blue,
    filecolor=magenta,      
    urlcolor=cyan,
    pdftitle={Overleaf Example},
    pdfpagemode=FullScreen,
    }

\newcommand{\refsec}[1]{Sect.~\ref{#1}}
\newcommand{\reftab}[1]{Table~\ref{#1}}
\newcommand{\refeq}[1]{Eq.~(\ref{#1})}
\newcommand{\reffig}[1]{Fig.~\ref{#1}}

\begin{document}

   \title{The \texttt{MARTINI} Platform (I): \\ Se I-X atomic calculation and expansion opacity for early-stage kilonova spectral analysis}

   \author{Matteo Bezmalinovich
          \inst{1}\fnmsep\inst{2}\fnmsep\inst{3}\thanks{matteo.bezmalinovich@inaf.it},
          Mattia Bulla
          \inst{2}\fnmsep\inst{4}\fnmsep\inst{5},
          Gediminas Gaigalas
          \inst{6},
          Diego Vescovi
          \inst{2}\fnmsep\inst{3},
          Matteo Canzari
          \inst{2} 
          \and \\
           Sergio Cristallo\inst{2}\fnmsep\inst{3}
          }

   \institute{Department of Physics, Sapienza University of Rome, P.le A. Moro 5, 00185-Rome, Italy
         \and
             National Institute for Astrophysics – Astronomical Observatory of Abruzzo, Via Maggini snc, 64100, Teramo, Italy
             \and
             National Institute for Nuclear Physics – Section of Perugia, Via A. Pascoli, Perugia, Italy
             \and
             Department of Physics and Earth Science, University of Ferrara, Via Saragat 1, I-44122 Ferrara, Italy
             \and
             National Institute for Nuclear Physics – Section of Ferrara, Via Saragat 1, I-44122 Ferrara, Italy
             \and
             Institute of Theoretical Physics and Astronomy, Vilnius University, Saulėtekio Ave. 3, Vilnius, Lithuania
             }

   \date{}

 
  \abstract
   {In the multi-messenger era, kilonovae represent key sites of r-process nucleosynthesis, making opacity estimation and spectral analysis crucial for constraining their composition. }
   {Since light r-process elements shape the early ($\sim\,0.5-1.5\,\mathrm{d}$) ejecta opacity, a detailed study of the selenium element with a focus on atomic data calculation, expansion opacity estimation and spectral analysis is presented.}
   {The selenium atomic data are calculated from Se I to Se X using the \texttt{GRASP2018} code. A systematic analysis and evaluation of their precision is performed through detailed comparison with the \texttt{NIST ASD}, and other works available in the literature. These atomic data are then used to estimate expansion opacity at different temperatures (e.g., $\mathrm{T}=5\,000\,\mathrm{K},\,10\,000\,\mathrm{K},\,20\,000\,\mathrm{K},\,100\,000\,\mathrm{K}$) and densities (e.g., $\rho = 10^{-13}\,\mathrm{g\,cm^{-3}},\,3\times10^{-12}\,\mathrm{g\,cm^{-3}}$). Spectral analysis has been performed with the Monte Carlo radiative transfer code \texttt{POSSIS} with a pre-computed opacity grid calculated with new densities and temperatures, ranging from -19.5 to -4.5 $\mathrm{g\,cm^{-3}}$ in log-scale and from $1\,000$ to $51\,000\,$ K, respectively. In the analysis, two scenarios are considered: one in which the opacity contribution comes from 100\% selenium ejecta, and another in which selenium contributes only partially to the total opacity ($\sim$ 10\% of the total mass).}
   {The selenium atomic calculations show a good agreement with \texttt{NIST ASD}, with accurate energy levels and transitions determined alongside atomic data for higher ionisation stages not fully covered by \texttt{NIST ASD}. The expansion opacities calculated with these new selenium data exhibit differences in comparison to existing literature works. Selenium spectral features can only be observed in the KN scenario consisting of 100\% selenium. When selenium accounts for about 10\% of the total KN mass, these features become undetectable. Finally, all selenium results are now available in the new open-source \texttt{MARTINI} platform dedicated to element nucleosynthesis.}
   {}

   \keywords{gravitational waves --
                atomic data --
                opacity --
                radiative transfer
               }

\titlerunning{The \texttt{MARTINI} Platform (I): Se I-X atomic calculation and expansion opacity}
\authorrunning{M. Bezmalinovich et al.}
\maketitle
%

\section{Introduction}
\label{sec:introduction}

Modern astronomy is characterized by an ever-growing stream of observational data. In the past (coming) years, the international astrophysics community has been (will be) inundated with spectroscopic data from numerous large-scale surveys (e.g., GAIA-ESO: \citealt{Randich2022}; GALAH: \citealt{DeSilva2015}; APOGEE: \citealt{Majewski2017}; 4MOST: \citealt{4most2019}; WEAVE: \citealt{weave2024}). Handling such a vast volume of data will require substantial theoretical efforts to accurately reconstruct the evolution of chemical abundances across time and space for the different components of our Galaxy, as well as in nearby galaxies.

Key ingredients in chemical evolution models are stellar yields. Some of them relate to stars that do not ignite carbon in their cores and end their evolution as white dwarfs, after passing through the Asymptotic Giant Branch (AGB) phase \citep[see, e.g.,][]{busso1999,straniero2006}. These stars are responsible for producing roughly half of the elemental abundances heavier than iron via the slow neutron-capture process \citep[s-process;][]{gallino1998}. 

The remaining half are produced by massive stars: during their pre-explosive phases via the weak s-process \citep[e.g.,][]{pignatari2010}, and during their final evolutionary stages—whether as single stars or members of binary systems—via the rapid neutron-capture process \citep[r-process; see, e.g.,][]{Cowan2021}. While the r-process yields of individual sites can vary due to the uncertainties in modelling the late evolutionary phases of stars, this does not substantially affect the global r-process abundance pattern. The s-process nucleosynthesis is anchored in laboratory measurements of all relevant reaction rates and half-lives, whereas the r-process relies largely on theoretical extrapolations for nuclei far from stability. For this reason, it is generally more robust to derive the solar r-process fraction as r=1-s \citep[see e.g.,][]{prantzos2020}, rather than inferring the s-process contribution from 1-r.

Interest in the r-process has surged in recent years following the gravitational wave event GW170817 \citep{Abbott2017} triggered by the coalescence of a binary neutron star (BNS) system. The GW170817 event was groundbreaking, as it marked the advent of multi-messenger astronomy \citep{Abbott2017,Corsi2024}. The GW detection was accompanied by electromagnetic counterparts. In particular, GW170817 included a short gamma-ray burst, identified as GRB170817A \citep{Abbott2017b, He2018, Lyman2018} and a kilonova (KN) emission stemming from the radioactive decay of r-process elements synthesised during the merger \citep[AT2017gfo;][]{Andreoni2017,Arcavi2017,Cowperthwaite2017,Drout2017,Evans2017,Kasliwal2017,McCully2017,pian2017,Smartt2017,SoaresSantos2017,Utsumi2017}. The KN provided the first direct evidence of in situ production of heavy elements via the r-process \citep[e.g.,][]{pian2017}. However, theoretical modelling of KNe requires a complex set of physical inputs, ranging from nucleosynthesis yields and thermalisation coefficients to line opacities of the individual chemical elements produced \citep[e.g.,][]{kasen2015,Metzger2019}.

\begin{figure}[htb!]
    \centering
    \includegraphics[width=1.0\linewidth]{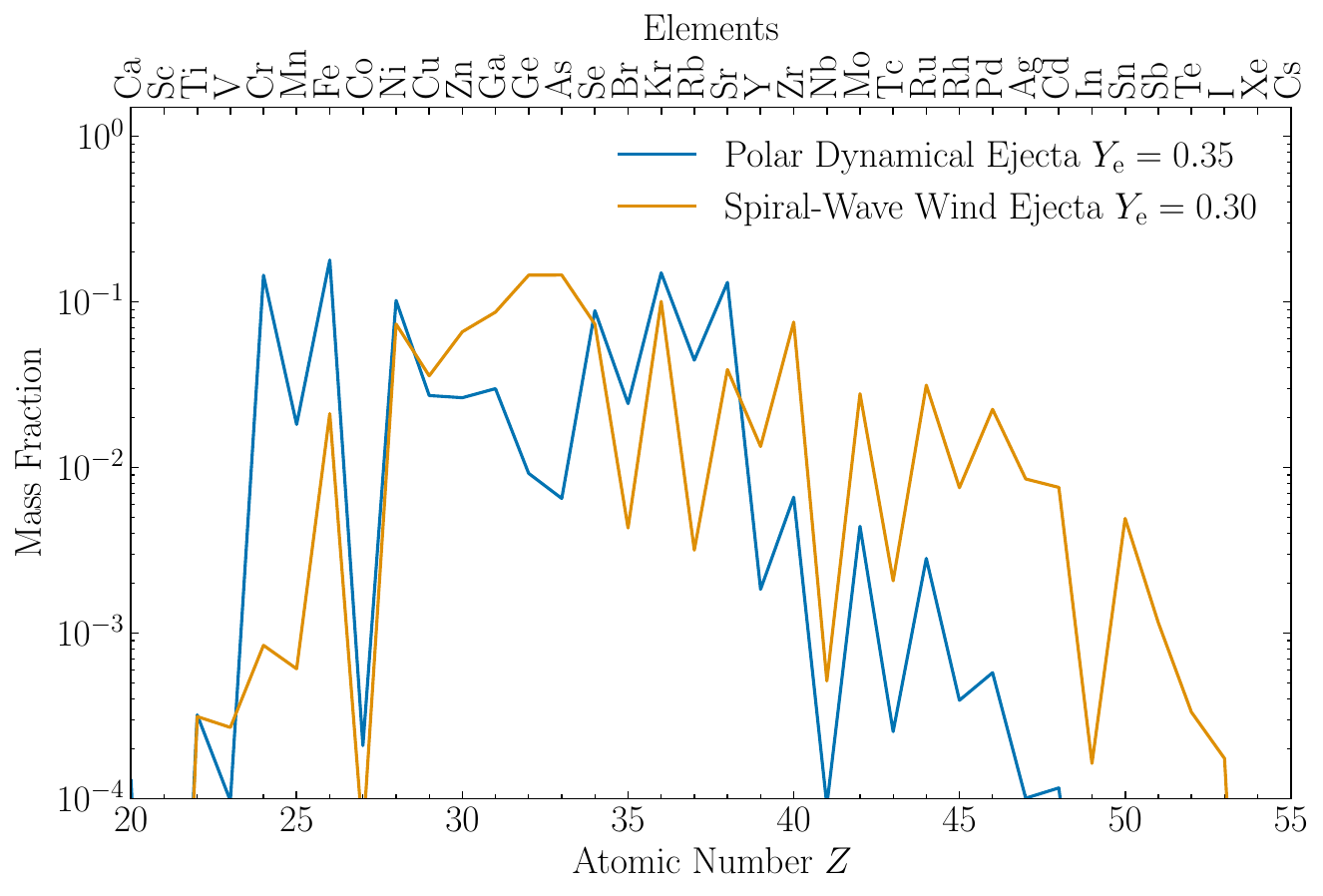}
    \caption{Abundances of light r-process elements for trajectories representative of the polar dynamical and spiral-wave wind, at 1.5 days after merger. For both trajectories, selenium accounts for about 10\% of the total mass.}
    \label{fig: Ye_trajectories}
\end{figure}

Most of the recent literature works as \citet{Banerjee2022, Banerjee2024}, \citet{Fontes2023}, \citet{Flrs2023}, \citet{CarvajalGallego2024}, \citet{Kato2024} have focused on lanthanides and actinides, which undoubtedly play a significant role in KN opacity. However, much less attention has been paid to non-lanthanide elements of the early stage ($\sim$ 0.5-1.5 days after the merger) evolution, especially at high-ionisation states. During this phase, light elements play a key role in shaping the early spectrum \citep{Kasen2013,Villar2017,Metzger2019}. Figure \ref{fig: Ye_trajectories} shows typical abundances of light r-process elements synthesized in the high-$Y_\mathrm{e}$ (high electron fraction) polar dynamical and spiral-wave ejecta of BNS mergers. 

These lanthanide-poor components, being the earliest and fastest ejecta, can dominate the blue kilonova emission during the first days after the merger \citep{Nedora2019,Perego2022}. The polar dynamical ($Y_\mathrm{e}$=0.35) and spiral-wave ($Y_\mathrm{e}$=0.30) trajectories shown here are representative examples of their respective ejecta types \citep{Vescovi2022}. In the polar dynamical case, nucleosynthesis produces Sr as the heaviest abundantly produced element, while the spiral-wave trajectory may also produce interesting elements such as Y, Mo, and Pd.

There are studies available in the literature that investigated strontium \citep{Watson2019, Perego2022, Domoto2022, Sneppen2024}, while less attention has been paid to other elements, such as yttrium \citep{Sneppen2023}, zirconium \citep{Domoto2022}, palladium, molybdenum, and selenium (Se) which are abundantly produced through the r-process. Moreover, their relatively simple atomic structure makes them easier to handle in both theoretical and experimental studies, providing valuable information on the opacity behaviour under KN conditions \citep{Mascali2022, Pidatella2021, Bezmalinovich2024}. 

To reduce the lack of data for highly ionised states of non-lanthanide elements, in this first article, we present a study focused on selenium (Se) I-X atomic data and expansion opacity calculation in the framework of early-stage kilonova. While our long-term goal is to include all elements and perform a fully consistent KN modelling, in this work we investigate only Se because it is one of the most abundant element produced by ejecta with high electron fractions ($Y_\mathrm{e}$ = 0.3-0.35) and mass fraction $X_\mathrm{Se}$=0.1 (\reffig{fig: Ye_trajectories}) in early-stage scenario. Moreover, Se was chosen as it is among the primary candidates for future measurement at PANDORA \citep{Mascali2022}, a magnetic trap facility currently under construction at the INFN - Laboratori Nazionali del Sud in Italy. Therefore, investigating its atomic properties and opacity is essential in view of future experiments and KN modelling.

Prompted by the need for these data from both theoretical and experimental communities, we have created a new online platform, \texttt{MARTINI}\footnote{https://martini.oa-abruzzo.inaf.it/}, designed to provide researchers with the inputs needed to simulate the chemical evolution of the Galaxy and to model KN light curves. 

The paper has been divided into different sections. In \refsec{sec: TheoryBackground}, we begin with a theoretical introduction to KN and expansion opacity. Moreover, a brief presentation of the \texttt{GRASP2018} (General Relativistic Atomic Structure Package) atomic code developed by \citet{FroeseFischer2019}, employed for the atomic calculation, and the KN code \texttt{POSSIS} \citep[POlarization Spectral Synthesis In Supernovae;][]{Bulla2019,Bulla2023} is presented. \refsec{sec: AtomicCalculation} is entirely dedicated to Se atomic calculations and to the computational strategy adopted. Specifically, starting from the available works of \citet{Tanaka2020}, \citet{Radit2022} and \citet{Kitovien2024}, we perform Se atomic calculations, with the aim of improving the accuracy of Se I–IV energy levels and transitions compared to the \texttt{NIST ASD}\footnote{https://www.nist.gov/pml/atomic-spectra-database} \citep{NISTASD}. We extend our approach to higher ionisation states, from Se V to Se X. These atomic results were then used to estimate the expansion opacity, which results are reported in \refsec{sec: ExpansionOpacity}. Finally, in \refsec{sec: SpectraAnalysis} we incorporated the opacity results into \texttt{POSSIS} to analyse and investigate KN spectral features.


\section{Theoretical background}
\label{sec: TheoryBackground}

\subsection{Kilonova physics}

KNe occur during the post-merger phase of a BNS or black hole - neutron star (BHNS) system shortly after mass ejection \citep{Radice2020}. They are quasi-thermal electromagnetic transients powered by the radioactive decay of heavy neutron-rich isotopes synthesized in the aftermath of BNS or BHNS mergers \citep{Kasen2017,Metzger2019}. The peak luminosity and colour of the transient depend on several factors (e.g., ejecta mass, velocity, composition) and can evolve with time. For AT2017gfo, the KN initially peaked in the near-ultraviolet/optical range, pushing to infrared emissions as soon as the ejected material expanded and cooled \citep[e.g.,][]{Chornock2017}

The study of KNe offers a unique opportunity to investigate the conditions under which heavy elements are synthesized and their subsequent distribution in the cosmos. The extreme neutron-rich environment of KN ejecta allows the efficient synthesis of heavy elements, such as gold, platinum \citep{Gillanders2021} and the majority of elements beyond the second r-process peak (A $\geq$ 130). 

However, due to several factors that are not yet well understood (e.g., EoS, the mass ratios of the neutron stars, nuclear physics rates and half lives, role of neutrinos, mass ejection processes, time scales), it is challenging to precisely determine the exact composition of the ejecta emitted from a BNS. An approach to addressing this issue involves analysing the observed spectra of the KN. 
Consequently, overcoming uncertainties in opacity calculations, which often arise from oversimplified atomic structure models \citep{Barnes2016,Kasen2019,Tanaka2020} is necessary to ensure precise spectral analysis and resolve between theoretical predictions and observational data for KNe.

\subsection{The expansion opacity}

KN opacity is dominated by bound-bound transitions. Following previous studies \citep{Kasen2013,Barnes2013, Tanaka2013, Tanaka2020}, we compute the expansion opacity using the formalism for bound-bound lines \citep{Sobolev1960, Castor1974, Karp1977}, combined with the formula of \citet{Eastman1993}.
During the post-merger phase of a BNS system, the ejecta expand at high velocities $v_\mathrm{ej}$ following the laws of homologous expansion with a timescale $t$ (this assumption may broke down in case of circumstellar material plasma through which the ejecta propagate). In this scenario, the spectral lines generated are broadened because of the dynamics of the ejecta.

The expression for calculating the opacity within a bin in an expanding medium is defined by the so-called bound–bound expansion opacity \citep{Eastman1993,Banerjee2020}:

\begin{equation}
    \kappa_{\mathrm{exp}}^{b-b}(\lambda)=\frac{1}{\rho c t_{\mathrm{exp}}}\sum_{l\in\Delta\lambda}\frac{\lambda_l}{\Delta\lambda}(1-e^{-\tau_l}) 
    \label{eq: ExpOpacity}
\end{equation}
where $t_{\mathrm{exp}}$ refers to the time elapsed since the material was ejected, $\rho$ is the density, $c$ the speed of light, $\lambda_l$ is the wavelength of the corresponding $l$ transition within the bin $\Delta\lambda$. $\tau_l$ represents the Sobolev optical depth, defined as 

\begin{equation}
    \tau_l = \frac{\pi e^2}{m_{\mathrm{e}} c}n_\ell f_l t_{\mathrm{exp}} \lambda_l 
    \label{eq: Tau_Sobolev} \,
\end{equation}
with $f_l$ the oscillator strength, $n_\ell$ the density population of $\ell$-level, $m_{\mathrm{e}}$ and $e$ the mass and charge of the electron, respectively. 

In the literature, there are two other approaches that are frequently used to calculate opacity. One is the line-binning \citep[e.g.,][]{Fontes2020}, while the other one is the line-by-line method \citep[e.g.,][]{Shingles2023}. The former consists of grouping spectral lines into wavelength bins and calculating the opacity within each bin, without including the ``expansion'' of the medium (typically for the expansion opacity). In the latter technique, the opacity is calculated for each individual line with no bin considerations. 

Calculating the expansion opacity requires detailed atomic data inputs. Following previous works \citep{Tanaka2020,Banerjee2022,Banerjee2024,Pognan2022}, we assume local thermodynamic equilibrium (LTE), which is a reasonable approximation in KN early-phase (< 2-3 days). Through the Saha equation, we determine the ionisation balance. This equation relates two adjacent stages of ionisation $i$ and $i+1$ and is defined as:
\begin{equation}
    \frac{n_{i+1}n_e}{n_i} = \frac{2U_{i+1}(T)}{U_{i}(T)}\frac{(2\pi m_ek_BT)^{3/2}}{h^3}e^{-\chi_i/k_BT} \ ,
    \label{eq: SahaEquation}
\end{equation}
where $k_\mathrm{B}$ is the Boltzmann constant, $n_\mathrm{e}$ is the electron number density, $h$ is the Planck constant, $\chi_i$ is the ionisation potential of the $i^{th}$ ionisation degree, and $U_i(T)$ is the partition function for the $i^{th}$ ionisation state defined as:
\begin{equation}
    U_i(T) = \sum_\ell g_{i,\ell} \ e^{-E_{i,\ell}/k_\mathrm{B}T} \ , 
    \label{eq: PartitionFunction}
\end{equation}
where $g_{i,\ell}$ is the statistical weight corresponding to the energy $E_{i,\ell}$ of the level $\ell$. Once the ionisation balance is determined, the Boltzmann distribution specifies the relative population of excited levels within a given ionisation stage,
\begin{equation}
    \frac{n_{i,\ell}}{n_i} = \frac{g_{i,\ell}}{U_i(T)} \exp\left(-\frac{E_{i,\ell}}{k_\mathrm{B} T}\right)\ ,
    \label{eq: Boltzmann_distribution}
\end{equation}
where $n_{i,\ell}$ is the density population of $\ell$-level with respect to the total population $n_i$ of the $i^{\mathrm{th}}$ ionisation stage.

As shown in \refeq{eq: ExpOpacity}, atomic calculations are essential to accurately compute the expansion opacity. In this context, we used the \texttt{GRASP2018} atomic code.

\subsection{The theory behind \texttt{GRASP2018}} 
\label{sec: theory_GRASP2018}

\texttt{GRASP2018} is a General Relativistic Atomic Structure Package developed by \citet{FroeseFischer2019} based on fully relativistic (four-component) multiconfiguration Dirac-Hartree-Fock (MCDHF) and relativistic configuration interaction (RCI) methods \citep{Grant2007,Fischer2016}, suitable for medium to heavy atomic systems. Starting from the eigenvalue equation $\hat{H}_{DC}\psi=E\psi$ with $\hat{H}_{DC}$ the Dirac–Coulomb Hamiltonian, defined as
\begin{equation}
    \hat{H}_{DC} = \sum^N_{j=1}\left( c\bm{\alpha}_j \cdot \bm{p}_j + (\beta_j -1)c^2 + V(r_j) \right) + \sum^N_{j<k} \frac{1}{r_{jk}}
\end{equation}
with $V$ the interaction potential between $N$ total electrons, $r_{jk}$ the distance of electron-electron repulsion between two distinguish electrons ($j,k$), \bm{$\alpha$} and $\beta$ the Dirac matrices and \bm{$p$} the momentum operator, the MCDHF method is applied to determine numerical representation of atomic state function (ASF) $\psi$ as 
\begin{equation}
    \psi(\gamma PJM_J) = \sum_r c_r\Phi(\gamma_rPJM_J) \ ,
    \label{eq: ASF}
\end{equation}
obtained by the linear combination of configuration state functions (CSF) $\Phi(\gamma_rPJM_J)$, eigenfunctions of the parity $P$, total angular momentum operators $J^2$ and its projection $M_J$, with the mixing coefficients $c_r$ \citep{Kitovien2024}.

\texttt{GRASP2018} performs a single run per element per ionisation state. It starts with a set of electron configurations (ECs) provided as input to determine a multireference (MR) set, followed by the estimation of the radial part of the wavefunction. In addition, angular integrations based on the second quantization formalism \citep{Gaigalas2001,Gaigalas2017} are included, together with Breit interaction and quantum electrodynamic (QED) corrections applied to the RCI computation. Finally, the energy levels and transitions are determined.

\begin{table*}[htb!]
    \caption{MultiReference sets of Se I-IV electron configuration with fixed core [Ar]3d$^{10}$. In each MR chosen, the same \{5s,5p,4d,4f\} orbital symmetry with the corresponding quantum numbers has been kept.}
    \label{tab: ECs}
    \centering
    \begin{tabular}{c c c c c c c c}
    \hline
    \hline
    \noalign{\smallskip}
    \multicolumn{2}{c}{Se I} & \multicolumn{2}{c}{Se II} & \multicolumn{2}{c}{Se III} & \multicolumn{2}{c}{Se IV} \\
    \noalign{\smallskip}
    \hline
    \noalign{\smallskip}
    $Even$ & $Odd$ & $Even$ & $Odd$ & $Even$ & $Odd$ & $Even$ & $Odd$ \\
    \noalign{\smallskip}
    \hline
    \hline
    \noalign{\smallskip}
    $4s^2 \,4p^4$  &  $4s^2 \,4p^3 \,4d$ & $4s^2 \,4p^2 \,4d$ &  $4s^2 \,4p^3$ & $4s^2 \,4p^2$ &  $4s^2 \,4p \,4d$ & $4s \,4p^2$ &  $4s^2 \,4p $  \\ 
    $4s^2 \,4p^2 \,4d^2$   & $4s^2 \,4p \,4d^3$  & $4s \,4p^4$ &  $4s \,4p^3 \,4d$ & $4p^4$ &  $4s^2 \,4p \,5s$ & $4s \,4p \,4f$ &  $4p^3$\\  
    $4s^2 \,4p^3 \,4f$   & $4s^2 \,4p^3 \,5s$  & $4s \,4p^3 \,4f$  &  $4s^2 \,4p^2 \,4f$ & $4s \,4d^3$ &  $4s^2 \,5s \,5p$ & $4s^2 \,4d$ & $4s^2 \,5p$ \\     
    $4s^2 \,4p^3 \,5p$ & $4s^2 \,4p^2 \,5s \,5p$  & $4s^2 \,4p^2 \,5s$  & $4s^2 \,4p \,5s^2$ & $4s^2 \,4d^2 $  & $4s \,4p^3$ & $4s \,4p \,5p$  & $4s^2 \,4f $ \\    
    $4s^2 \,4p^2 \,5s^2$  & $4s^2 \,4p^2 \,4d \,4f$  & $4s^2 \,4p \,5s \,5p$ & $4s^2 \,4p^2 \,5p$ &  $4s^2 \,4p \,5p$   & $4s \,4p^2 \,4f$ & $4s^2 \,5s$ & $4s \,4p \,5s$ \\      
    & & & & $4s^2 \,4p \,4f$  & & & $4s \,4p \,4d$\\  
    & & & & $4s \,4p^2 \,5s$    &  & &\\
    \noalign{\smallskip}
    \hline
    \hline
    \end{tabular}
\end{table*}

\subsection{KN spectra modelling: \texttt{POSSIS}} 
\label{sec: theory_POSSIS}

Once the atomic calculations are performed, it is possible to estimate the expansion opacity, and the results are then inserted as input into \texttt{POSSIS}  \citep{Bulla2019,Bulla2023}, a three-dimensional Monte Carlo radiative transfer code that can model radiation transport in expanding media like the ejecta of supernovae and KNe. Unlike other sophisticated radiative transfer codes available in the literature \citep[e.g.,][]{Hoflich1993,Blinnikov1998,Hauschildt1998,Utrobin2004,Dessart2005,Kasen2006,Kromer2009,Tanaka2013,Wollaeger2014,Kerzendorf2014}, \texttt{POSSIS} uses pre-computed atomic opacity tables as input, which can significantly speed up the calculation time. The code includes wavelength- and time-dependent opacities and generates angular-dependent predictions of spectra, light curves, and polarization for both simplified and hydrodynamical explosion models. We report below the key aspects of the code's structure and focus particularly on the latest version of the code \citep{Bulla2023} that is used in this work. More details can be found and studied accurately in \citet{Bulla2019} and \cite{Bulla2023}.

First, the ejecta are described by a 3D Cartesian grid specifying for each cell $i^{\mathrm{th}}$ spatial coordinates, density $\rho_{i,0}$ and electron fraction $Y_{{\mathrm{e}},i}$ at a reference time $t_0$. The time-dependence is modelled assuming homologous expansion, i.e. the velocity and mass of each cell is constant and therefore radial coordinates and densities scale with time $t$ as $r_i\propto t$ and $\rho_i\propto t^{-3}$, respectively. We assume $Y_{{\mathrm{e}},i}$ to remain constant within each cell, i.e. the composition is fixed during the simulation.

Key properties controlling the energy budget and setting the temperature profile of the ejecta are the nuclear heating rates $\dot{\epsilon}(t)$ and the thermalisation efficiencies $\epsilon_{\mathrm{th}}$. Nuclear heating rates -- providing the amount of energy per unit time per unit mass produced by the radioactive decay of r-process nuclei -- are taken from \cite{Rosswog2022}. In this framework, the electron fraction $Y_\mathrm{e}$ and the expansion velocity $v_\mathrm{exp}$ enter only as input parameters required to select the appropriate heating-rate prescription and are therefore fixed to their initial values within each cell. Although $Y_\mathrm{e}$ and the chemical composition may evolve with time, this evolution is not explicitly followed, as its effect is already captured by the time dependence of the tabulated heating rates. Instead, the expansion velocity is computed at the start of the simulation as a function of the cell mass and the cell width in velocity space, which remain constant under homologous expansion. As a result, for each cell the heating rate depends solely on time. The thermalisation efficiencies -- controlling what fraction of the heating rates thermalise within the ejecta and is available to power the kilonova -- are computed within the code following standard prescriptions \citep{Barnes2016,Wollaeger2018}.

Assuming perfect coupling between matter and radiation, heating rates and thermalisation efficiencies are then used to initialize the temperature of the ejecta. Specifically, the mean intensity of the radiation field in each cell is computed as
\begin{equation}
<J_i>=\frac{c}{4\pi}\int_0^{t_{\mathrm{s}}}\rho_i(t_{\mathrm{exp}})\,\dot{\epsilon_i}(t_{\mathrm{exp}})\,\epsilon_{\mathrm{th,i}}\,\bigg(\frac{t_{\mathrm{exp}}}{t_{\mathrm{s}}}\bigg)\,{\mathrm{d}}t_{\mathrm{exp}}  
\end{equation}
and the ejecta temperature is set equal to the radiation temperature $T_i$ estimated via the Stefan-Boltzmann law:
\begin{equation}
T_i=\bigg(\frac{\pi\,<J_i>}{\sigma}\bigg)^{1/4}~~~~~.
\label{eq:SB}
\end{equation}
In the two equations, the $t_{\mathrm{exp}}$/$t_{\mathrm{s}}$ term accounts for adiabatic energy losses from a given time $t_{\mathrm{exp}}<t_{\mathrm{s}}$ to the starting time of the simulation $t_{\mathrm{s}}$, and $\sigma$ is the Stefan-Boltzmann constant. During the simulation, the temperature is updated at the end of every time-step using in Eq.~\ref{eq:SB} Monte Carlo estimators \citep{Mazzali1993,Lucy2003} for the mean intensity,
\begin{equation}
<J_i>=\frac{1}{4\pi\Delta t\,V}\sum{e_{\mathrm{cmf}}\Delta l}  
\label{eq:MCest}
\end{equation}
where $V$ is the cell volume and the sum is performed over all the Monte Carlo photon packets with comoving-frame energy $e_{\mathrm{cmf}}$ travelling a path length $\Delta l$ through the given cell at the given time-step with duration $\Delta t$.


Once the grid structure is set, a number $N_{\mathrm{ph}}$ of identical and indivisible Monte Carlo photon packets \citep{Abbott1985} are created and spread evenly across the different time-steps. Monte Carlo photons activated at a given time-step are assigned initial positions $\boldsymbol{x}$, initial directions $\boldsymbol{n}$, energies $e$, frequencies $\nu$ and normalized Stokes vectors $\boldsymbol{s}$. The initial energy of all packets is the same \citep{Abbott1985} and computed by  evenly splitting the energy from the radioactive decay of r-process nuclei thermalised within the ejecta \citep{Wollaeger2018,Rosswog2022}. The initial frequency $\nu$ is chosen by sampling the thermal emissivity
\begin{equation}
    S(\nu) = B(\nu,T) \cdot \kappa_{\mathrm{tot}} (\nu) \ ,
    \label{eq: ThermalEmissivity}
\end{equation}
where $\kappa_{\mathrm{tot}}(\nu)$ is the total opacity and $B(\nu,T)$ is the Planck function at temperature $T$. 

The propagation of Monte Carlo packets is then controlled by the opacities of the expanding medium. In particular, \texttt{POSSIS} can take as input tables specifying electron-scattering and bound-bound opacities as a function of wavelength, time and local ejecta properties like density, temperature and electron fraction\footnote{
We note that using different inputs for heating rates and opacities and decoupling them through their dependence on $Y_{\mathrm{e}}$ may break self-consistency. In the future, we plan to carry out simulations where heating rates and opacities are computed self-consistently for specific compositions of the ejecta.} . The output flux and polarization spectra as a function of time are then constructed `on the fly' using estimators \citep{Kerzendorf2014,Bulla2015}, a technique that leads to a significant reduction in numerical noise compared to the binning of escaping packets commonly used in Monte Carlo radiative transfer simulations.


\section{Atomic calculation}
\label{sec: AtomicCalculation}

\subsection{\texttt{GRASP2018} calculation strategy} 
\label{sec: GRASP2018_strategy}

We start by focusing on the first four degrees I-IV of Se (i.e., Se I, Se II, Se III, and Se IV). A MR set of ECs must be defined to determine the CSFs for the ASF, as seen in \refeq{eq: ASF}. \reftab{tab: ECs} report the MR sets of the ECs selected for the calculation of Se I-IV. 

The ECs are divided by parity, and their relative wavefunctions are optimized separately according to even or odd states. The choice of ECs was guided by references such as the \texttt{NIST ASD} \citep{NISTASD} and articles available in the literature by \citet{Tanaka2020}, \citet{Radit2022}, and \citet{Kitovien2024}. Moreover, we perform our calculations by respecting the same orbital symmetry between even and odd states for all the degrees of ionisation, ensuring convergence within \texttt{GRASP2018}. At odds with the work of \citet{Tanaka2020}, our calculation includes \textit{f}-shell orbitals for Se I to IV. On the one hand, this approach allows us to highlight possible differences between the two works and, on the other hand, allows us to enable a well-founded comparison with \texttt{NIST ASD} and the two works \citet{Radit2022} and \citet{Kitovien2024}, both of which provide \textit{f}-shell ECs. To calculate the energy levels and the transitions, we include single-double (SD) excitation applied to the valence shell, while keeping the core [Ar]3d$^{10}$ and the 4s$^2$ shells frozen during the whole calculation. For each ionisation stage, we apply the same computational strategy.
\newline
Once Se I–IV are complete, the procedure is extended to Se V–X, using the corresponding ECs reported in \reftab{tab: ECs_V-X}. 

\begin{table}[htb!]
\caption{MR sets of Se V-X with fixed core [Ar] for Se V, VI and VII, and [Ne] for Se VIII-X. In Se V the same \{5s,5p,4d,4f\} orbital symmetry was kept, while in VI and VII it was broken to ensure convergence. In the cases of Se VIII-X, the new orbital symmetry is \{5s,5p,4d\}}
    \label{tab: ECs_V-X}
    \centering
    \begin{tabular}{c c c}
    \hline
    \hline
    \noalign{\smallskip}
     & $Even$ & $Odd$  \\
     \noalign{\smallskip}
    \hline
    \hline
    \noalign{\smallskip}
    \multirow{6}{*}{Se V} & $3d^{10} \,4s^2$  &  $3d^{10} \,4s \,4p$ \\
      & $3d^{10} \,4p^2$  &  $3d^{10} \,4s \,5p$ \\
      & $3d^{10} \,4s \,4d$  &  $3d^{10} \,4s \,4f$ \\
      & $3d^{10} \,4s \,5s$  &  $3d^{10} \,4p \,4d$ \\
      & $3d^{10} \,4p \,4f$  &  $3d^{10} \,4p \,5s$ \\
      & $3d^{10} \,4p \,5p$  &   \\
      \noalign{\smallskip}
       \hline
       \noalign{\smallskip}
    \multirow{3}{*}{Se VI} & $3d^{10} \,4s$ &  $3d^{10} \,4p$ \\
      & $3d^{10} \,4d$ &  $3d^{10} \,5p$ \\
      & $3d^{10} \,5s$ &  $3d^{10} \,4f$ \\
      \noalign{\smallskip}
      \hline
      \noalign{\smallskip}
    \multirow{4}{*}{Se VII} & $3d^{10}$ &  $3d^{9} \,4p$ \\
      & $3d^{9} \,4d$ &  $3d^{9} \,5p$ \\
      & $3d^{9} \,4s$ &  $3d^{9} \,4f$ \\
      & $3d^{9} \,5s$ &   \\
      \noalign{\smallskip}
      \hline
      \noalign{\smallskip}
    \multirow{2}{*}{Se VIII} & $3s^2 \,3p^6 \,3d^9$ &  $3s^2 \,3p^5 \,3d^{10}$ \\
      & $3s^2 \,3p^6 \,3d^8 \,4s$ &  $3s^2 \,3p^6 \,3d^8 \,4p$ \\
      \noalign{\smallskip}
      \hline
      \noalign{\smallskip}
    \multirow{2}{*}{Se IX} & $3s^2 \,3p^6 \,3d^8$ &  $3s^2 \,3p^5 \,3d^9$ \\
      & $3s^2 \,3p^6 \,3d^7 \,4s$ &  $3s^2 \,3p^6 \,3d^7 \,4p$ \\
      \noalign{\smallskip}
      \hline
      \noalign{\smallskip}
    \multirow{2}{*}{Se X} & $3s^2 \,3p^6 \,3d^7$ &  $3s^2 \,3p^5 \,3d^8$ \\
      & $3s^2 \,3p^6 \,3d^6 \,4s$ &  $3s^2 \,3p^6 \,3d^6 \,4p$ \\
      \noalign{\smallskip}
      \hline
      \hline
    \end{tabular}
\end{table}

In particular, Se V is computed by adopting the same calculation strategy as in the lower ionisation stages. Starting from Se VI, a slight modification to this approach became necessary. Specifically, in the cases of Se VI and VII, the orbital symmetry involving the subshells \{5s, 5p, 4d, 4f\} is not preserved between the even and odd parity states, as this adjustment is required to ensure the convergence of the code. Beyond Se VII, the \textit{f}-shell is no longer included, because the valence electrons in these higher ionisation states primarily occupy the \textit{s}, \textit{p} and \textit{d}-shells. Given the high temperature ($\gtrsim$ 10 000 K) of the medium, electrons are more likely to be ionised rather than excited to the \textit{f}-shell. Furthermore, from Se VIII, the [Ne] core is treated as frozen throughout the entire calculation, including a new orbital symmetry \{5s, 5p, 4d\} between even and odd states. Finally, the case of Se IX and X is complicated by the lack of guidance from the \texttt{NIST ASD}, which does not provide information on possible ECs. To address this gap, a set of ECs is independently defined, with a focus on improving the transition probabilities and improving the accuracy of the ASF, thus strengthening the final opacity calculations. 
\newline
\begin{figure*}[htb!]
    \centering
    \includegraphics[width=0.9\linewidth]{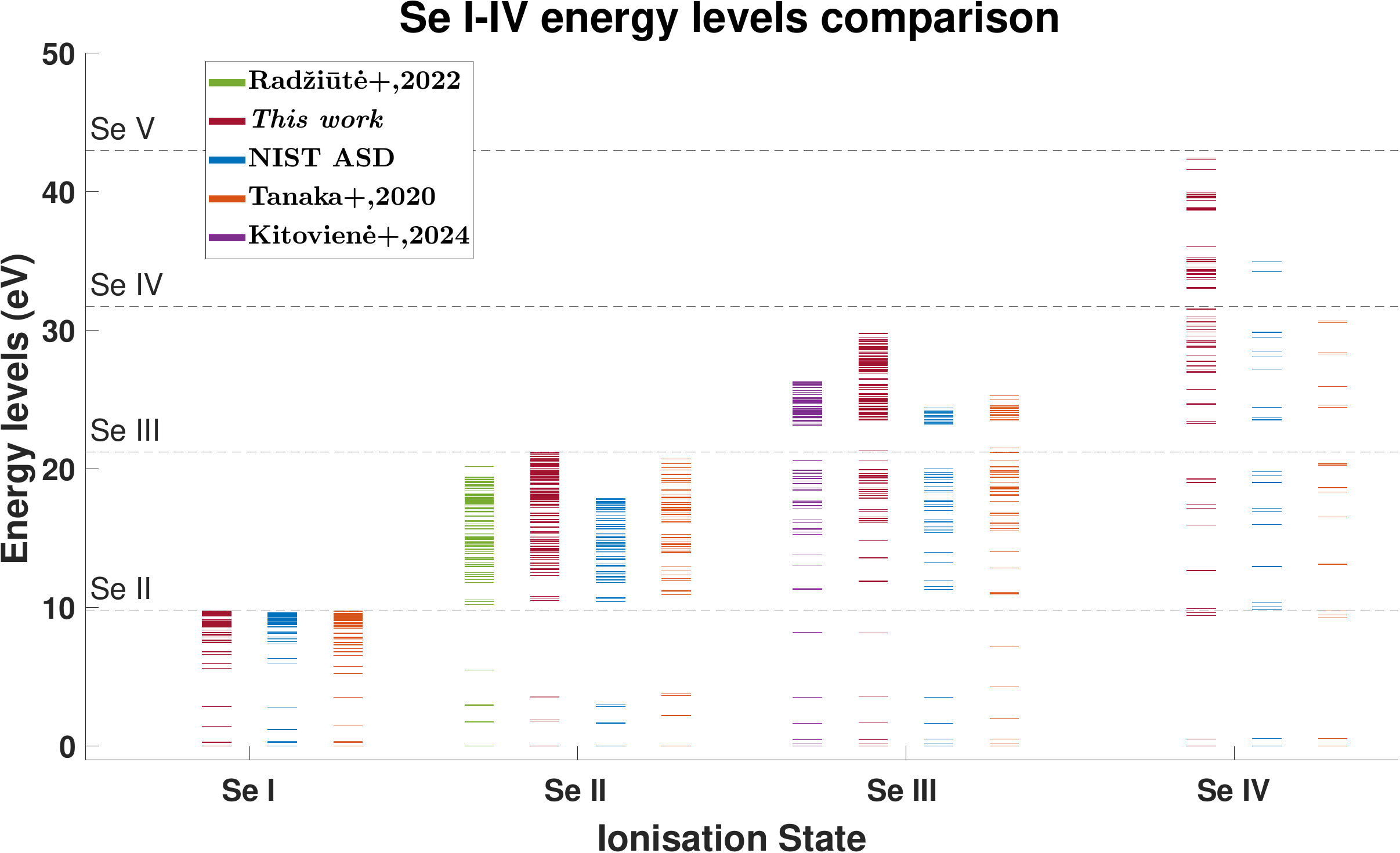}
    \caption{Se I-IV energy levels of the four works compared with the \texttt{NIST ASD}. In the analysis we considered all energy levels below ionisation for each state. The dashed lines represent the ionisation threshold for each ionisation degree.}
    \label{fig: SeI-IV_energy_levels}
\end{figure*}
In \texttt{GRASP2018}, RCI correlations with MR set are included taking into account a set of virtual orbitals with a step-by-step procedure through different layers. In our calculation, we assume three different layers, starting from the initial L$_1$ to the final layer L$_3$.

\begin{table}[htb!]
\caption{The list of different layers used for the RCI is as follows.}
\label{tab: SOL}
\centering
\begin{tabular}{c|c c|c c|c c}
\hline
    & \multicolumn{2}{c|}{Se I-V} & \multicolumn{2}{c|}{Se VI-VII} & \multicolumn{2}{c}{Se VIII-X} \\
    \hline
    Layers & $Even$ & $Odd$ & $Even$ & $Odd$ & $Even$ & $Odd$  \\
    \hline
    \hline
    L$_1$ & \multicolumn{2}{c|}{\{6s,6p,5d,5f\}}  & \{6s,5d\} & \{6p,5f\} & \multicolumn{2}{c}{\{6s,6p,5d\}}  \\
    L$_2$ & \multicolumn{2}{c|}{\{7s,7p,6d,6f\}}  & \{7s,6d\} & \{7p,6f\} & \multicolumn{2}{c}{\{7s,7p,6d\}}  \\
    L$_3$ & \multicolumn{2}{c|}{\{8s,8p,7d,7f\}}  & \{8s,7d\} & \{8p,7f\} & \multicolumn{2}{c}{\{8s,8p,7d\}}  \\
\hline
\end{tabular}
\end{table}

As shown in \reftab{tab: SOL}, the adopted layer structure varies across the Se I to Se X calculations. This is necessary to ensure convergence of the atomic calculations, and to optimise the final energy levels and atomic transitions. In this regard, to correctly estimate the accuracy of the calculated energy levels, we compute mean absolute differences between energy levels, following a methodology presented in \citet{Radit2022}. In detail, this approach consists in computing the mean absolute differences and the mean relative differences between the predicted energy levels and those reported in the \texttt{NIST ASD} database, which serves as the reference. The equations used for these calculations are defined as follows:

\begin{equation}
    | \overline{\Delta E} | = \frac{\sum|\Delta E|}{N} = \frac{\sum|E^{\texttt{NIST}} - E^*|}{N}
    \label{eq: MeanAbsoluteError}
\end{equation}

\begin{equation}
    \frac{|\overline{\Delta E}|}{E} = \frac{\sum|\Delta E|/E}{N} = \frac{\sum|E^{\texttt{NIST}} - E^*|/E^{\texttt{NIST}}}{N} 
    \label{eq: MeanRelativeError}
\end{equation}

\begin{table*}[htb!]
    \caption{Se I-IV error estimation of the energy levels compared to \texttt{NIST ASD}. $\Delta$E$_{\mathrm{abs}}$ is referred to the mean absolute difference, while $\Delta$E$_{\mathrm{rel}}$ to the mean relative difference in percentage. The comparison with \texttt{NIST ASD} was made also with other available works in literature.}
    \label{tab: ErrorEstimation_Energy}
    \centering
    \begin{tabular}{c|c c|c c|c c|c c}
    \hline
    & \multicolumn{2}{c}{\textbf{\textit{This work}}} & \multicolumn{2}{c}{\textbf{M. Tanaka et al. (2020)}} & \multicolumn{2}{c}{\textbf{L. Radžiūtė et al. (2022)}} & \multicolumn{2}{c}{\textbf{L. Kitovienė et al. (2024)}}\\
    \hline
       Element  & $\Delta$E$_{\mathrm{abs}}$ & $\Delta$E$_{\mathrm{rel}}$ (\%)  & $\Delta$E$_{\mathrm{abs}}$ & $\Delta$E$_{\mathrm{rel}}$ (\%) & $\Delta$E$_{\mathrm{abs}}$ & $\Delta$E$_{\mathrm{rel}}$ (\%) & $\Delta$E$_{\mathrm{abs}}$ & $\Delta$E$_{\mathrm{rel}}$ (\%) \\
    \hline
    \hline
        Se I & 0.3481 eV & 5.53 & 0.5375 eV & 8.43 & / & / & / & / \\
        Se II & 0.3306 eV & 4.28 & 0.3387 eV & 5.96 & 0.1691 eV & 1.64 & / & / \\
        Se III & 0.4050 eV & 3.15 & 0.4484 eV & 4.20 & / & / & 0.1038 eV & 0.98 \\
        Se IV & 0.5862 eV & 3.66 & 1.5831 eV & 8.25 & / & / & / & / \\

    \hline
    \end{tabular}
\end{table*}

where $E^*$ denotes the energy level under investigation, to be compared with the corresponding value of the \texttt{NIST ASD} database, in a total of $N$ energy levels in common between the references \citep[i.e., ][]{Tanaka2020,Radit2022,Kitovien2024}. Identification of the levels obtained in the GRASP package is usually performed in the jj-coupling scheme. Therefore, in order to compare the obtained levels with the NIST database, where levels are usually identified in LS-coupling, it is necessary to perform a transformation of the atomic state function from jj-coupling to LS-coupling \citep{https://doi.org/10.48550/arxiv.physics/0405060,Gaigalas2003}. This transformation is performed by the jj2lsj program \citep{Gaigalas2004,Gaigalas2017}, which is integrated into the GRASP package \citep{FroeseFischer2019}. This identification of levels by LS-coupling and comparison with the NIST database is approved by the atomic theory community.

To broaden the analysis and achieve a complete and consistent evaluation, in addition to our Se results, we include available atomic data from the literature. We consider atomic results for Se II and Se III performed with \texttt{GRASP2018} and published by \citet{Radit2022} and \citet{Kitovien2024}, respectively, and Se I-IV data published by \citet{Tanaka2020} and available from the online database\footnote{http://dpc.nifs.ac.jp/DB/Opacity-Database/} Japan–Lithuania \citep{Lith-Jap} calculated using the HULLAC atomic code \citep{BarShalom2001}.

Finally, we adopt the same strategy and the quantitative and qualitative evaluation (QQE) method described in the work of \citet{Kitovien2024} to evaluate the accuracy of the calculated transitions. It consists of evaluating the derived transitions in different classes, based on the ASF accuracy of the wavefunction introduced in \refeq{eq: ASF}. The classes are as follows: AA, A+, A, B+, B, C+, C, D+, D, and E. Class E indicates a transition with low ASF accuracy, while AA represents the class with the highest.

\subsection{Se atomic calculation results}
\label{sec: SeResults}
Since our goal is to calculate the expansion opacity arising from bound-bound transitions, we restrict our analysis to bound-state energy levels only. We first report the results inherent in the energy levels of Se I-IV and Se V-X. Secondly, the resulting atomic transitions of all ionisation stages are presented.

\reffig{fig: SeI-IV_energy_levels} shows a comparison of the energy levels predicted by this work, and the published data of \citet{Tanaka2020}, \citet{Radit2022}, \citet{Kitovien2024}. The results of each work are compared with the \texttt{NIST ASD} and the error estimates are provided in \reftab{tab: ErrorEstimation_Energy}. 

\begin{figure}[htbp]
    \centering
    \includegraphics[width=1.0\linewidth]{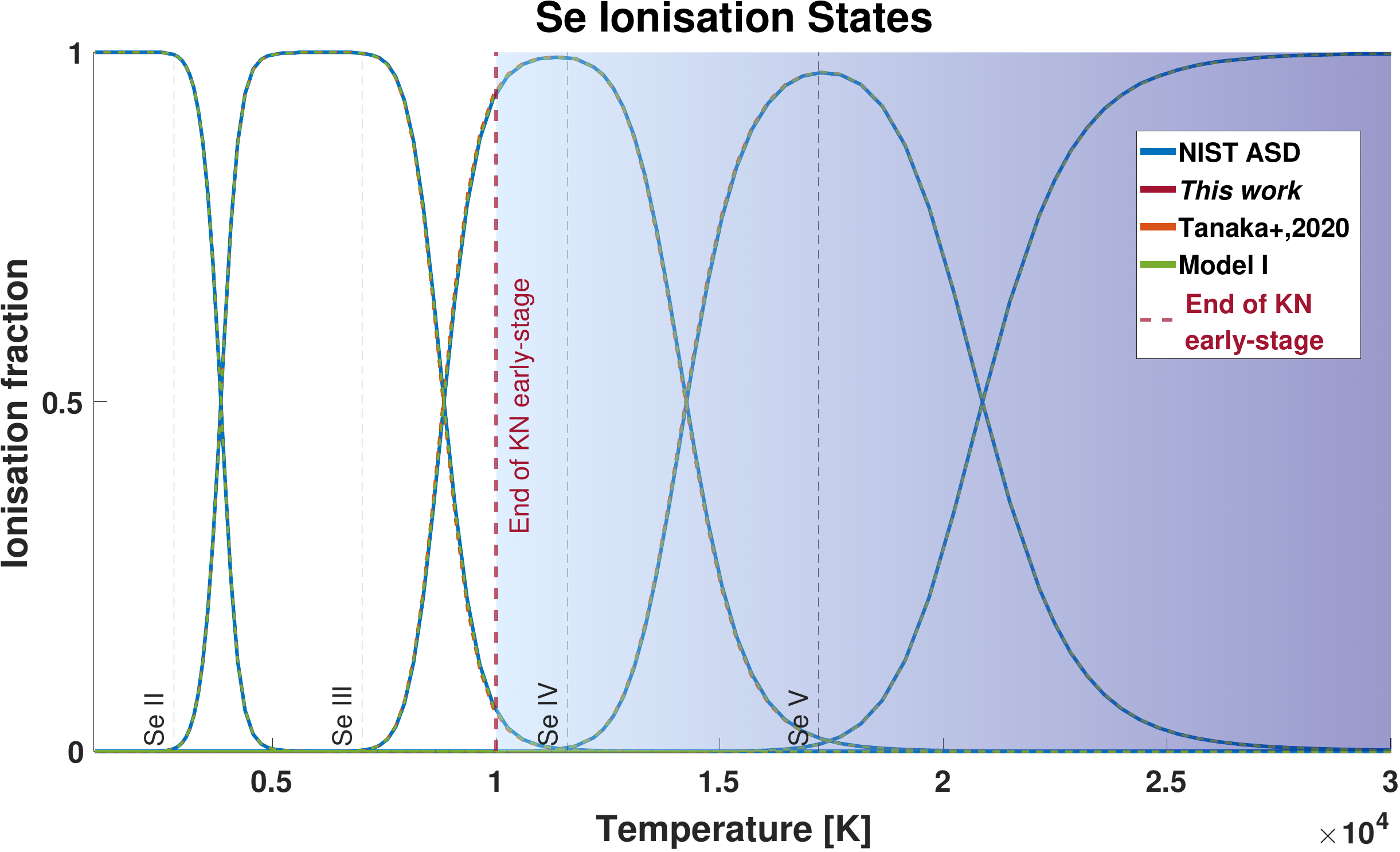}
    \caption{Plot of Se ionisation fraction as a function of the temperature. Vertical dashed lines indicates the temperature at which a new ionisation state is rising. The blue panel represents the range of temperatures of interest for the early-stage KN ($\gtrsim$ 10 000 K)}.
    \label{fig: Se_ionisation_fraction}
\end{figure}

\begin{figure*}[htb!]
    \centering
    \includegraphics[width=0.9\linewidth]{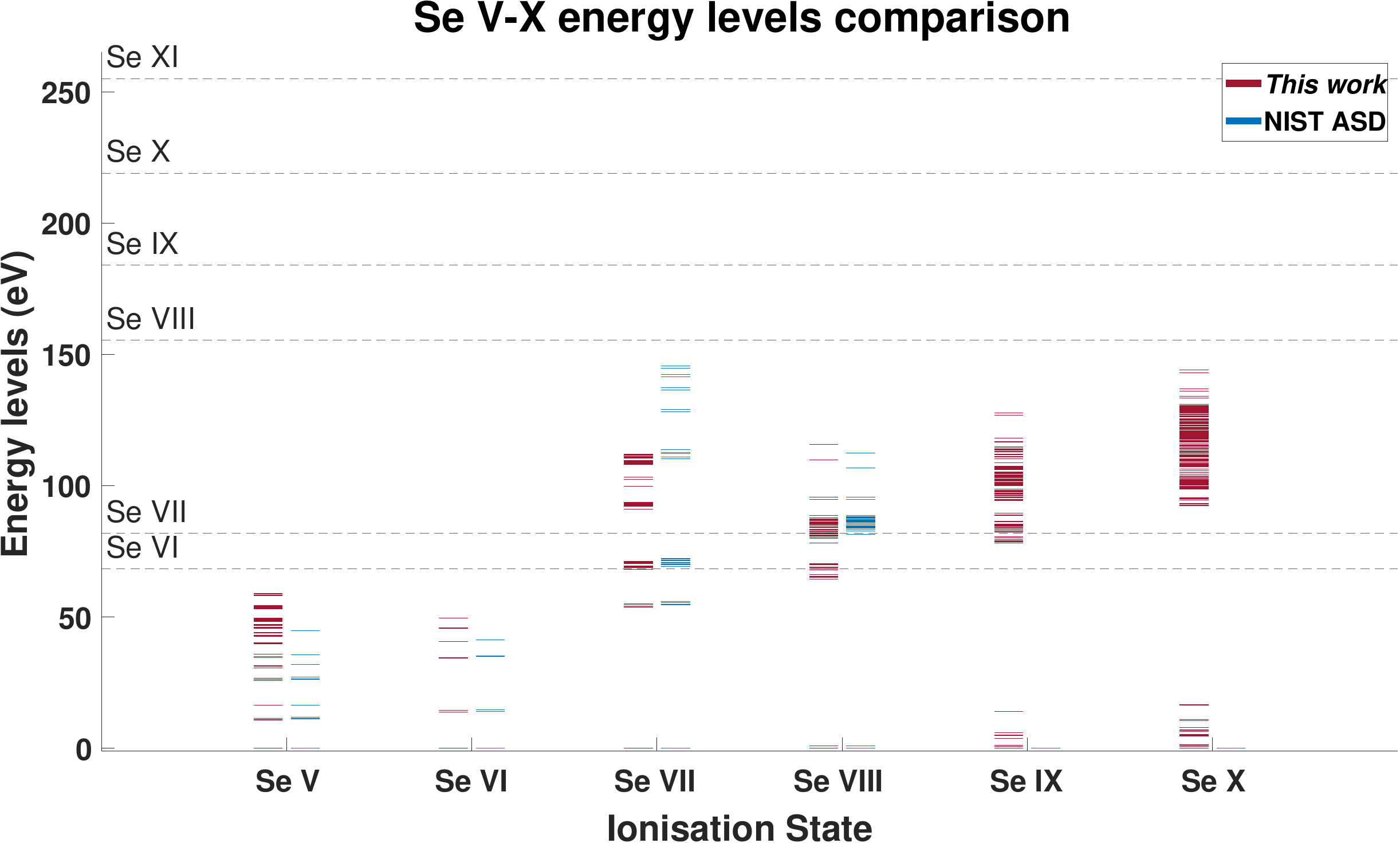}
    \caption{Se V-X energy levels of our \texttt{GRASP2018} atomic calculations compared with the \texttt{NIST ASD}. In the analysis we considered all energy levels below ionisation for each state. The dashed lines represent the ionisation threshold for each ionisation degree.}
    \label{fig: energyBeyond_SeIV}
\end{figure*}

Our Se I-IV calculations show good agreement with the \texttt{NIST ASD} and appear to be more accurate than the data reported in \citet{Tanaka2020} (see \reffig{fig: SeI-IV_energy_levels} and \reftab{tab: ErrorEstimation_Energy}). These differences in precision and number of energy levels, compared to \citet{Tanaka2020}, may be attributed to several factors, like the computational strategy (e.g., the inclusion of \textit{f}-shells in the ECs and RCI method adopted; see \reftab{tab: ECs} and \reftab{tab: SOL}) or the different atomic code used in the calculation.
Specifically, the accuracy of our Se I and Se IV energy levels improves by approximately 3\% and 5\%, respectively, while Se II and Se III show improvements around 2\% and 1\%, respectively. In this respect, we stress that the uncertainty of the Se IV ionisation state is more than halved with respect to previous available publications. Furthermore, our findings for Se II and Se III demonstrate that the works of \citet{Radit2022} and \citet{Kitovien2024} agree better with the \texttt{NIST ASD}, with results approximately two times more accurate than ours. Since the assumptions made by the two models are the same as those used by us (e.g., same atomic code and same inclusion of $f$-shell in ECs), these differences in the results highlight the need for a dedicated refinement of the choice of ECs and the RCI layer method in our \texttt{GRASP2018} calculations for each ionisation stages. However, achieving this level of precision on the results requires a huge effort and a very sophisticated computational procedure that takes months or even years to complete. In this context, our method sought a solution that is certainly less accurate but still provides a good level of precision of the data, allowing calculations to be feasible in a short time ($\sim$ from a few weeks to a few months). Moreover, the goal of our work is to provide reliable atomic data for several atomic elements, starting from the Se calculations presented in this work. Although not precisely accurate as in the works \citet{Radit2022} and \citet{Kitovien2024}, the computational strategy adopted by us demonstrates that the obtained Se results are in good agreement with the \texttt{NIST ASD} reference, and better reproduce accurate energy levels of Se I-IV than existing works in literature \citep[e.g.,][]{Tanaka2020}.

Further confirmation of the accuracy of the energy levels is obtained from the ionisation fraction plot. \reffig{fig: Se_ionisation_fraction} reports the ionisation fraction of our Se I-IV as a function of the temperature. This plot is obtained by applying the Saha-Boltzmann equations (see Eqs. \ref{eq: SahaEquation} and \ref{eq: Boltzmann_distribution}) on the atomic levels presented in \reftab{tab: ErrorEstimation_Energy} to correctly estimate the ionisation states and the population distribution.

In the same plot, we compare our Se I-IV results with the ionisation fraction calculated for other two models: one, namely ``Model I'', formed by the inclusion of \citet{Radit2022} Se II and \citet{Kitovien2024} Se III to our Se I and Se IV to create a model with the highest accuracy of Se data compared to the \texttt{NIST}. The other model considers Se I-IV atomic data of \citet{Tanaka2020}. First of all, it can be seen that the level population of our Se I-IV overlaps with the other two models, meaning that our Se I-IV energy levels contributing to the partition function have been estimated correctly. Moreover, from \reffig{fig: Se_ionisation_fraction}, it can be evinced that Se I and Se II do not impact as much as Se III and Se IV in the range of interest for early-stage KN. Rather, the ionisation fraction plot suggests that Se V should be taken into account for a more accurate KN modelling at these temperatures.

\begin{figure*}[htb!]
    \centering
    \includegraphics[width=\linewidth]{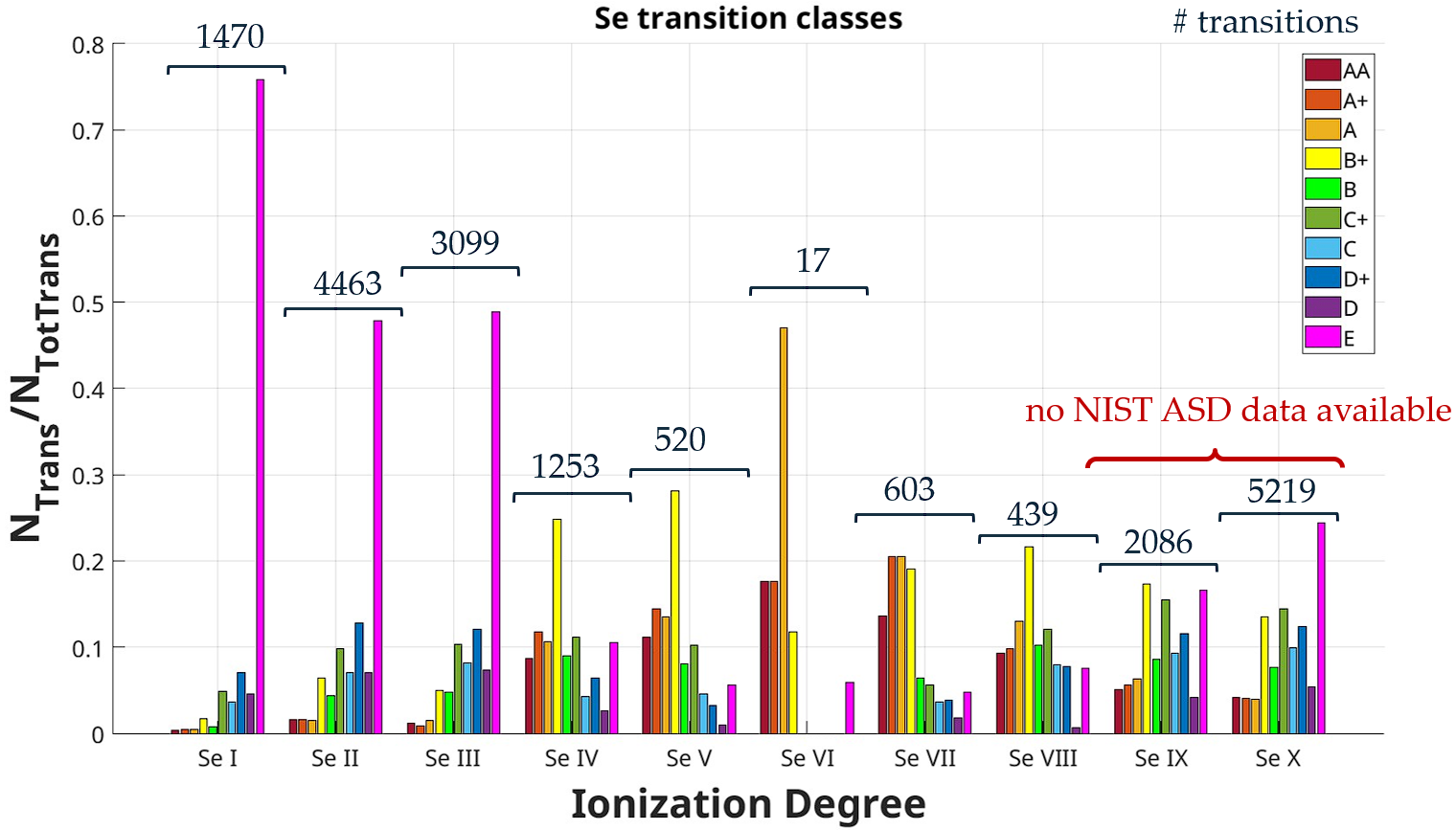}
    \caption{Se I-X Transition classes of \texttt{GRASP2018} calculation. In the legend, ``AA'' denotes transitions computed with highly accurate wave functions, whereas ``E'' indicates the least accurate ones.}
    \label{fig: Full_Transition}
\end{figure*}

Beyond Se IV, only theoretical models of \cite{Banerjee2024} are available in the literature, but without published or accessible data to be used for the analysis. Therefore, the evaluation of our results is based only on a comparison with the \texttt{NIST ASD} database. \reffig{fig: energyBeyond_SeIV} presents the calculated energy levels from Se V to Se X, with their respective error estimates listed in \reftab{tab: ErrorEstimation_Energy_SeV-X}.

\begin{table}[htb!]
    \caption{Error estimation of the energy levels between our \texttt{GRASP2018} atomic calculation and \texttt{NIST ASD}. $\Delta$E$_{abs}$ is referred to the mean absolute difference, while $\Delta$E$_{rel}$ to the mean relative difference in percentage. For Se IX and Se X there are no \texttt{NIST ASD} data available, hence it was not possible to estimate the errors.}
    \label{tab: ErrorEstimation_Energy_SeV-X}
        \centering
    \begin{tabular}{c|c c}
    \hline
    & \multicolumn{2}{c}{\textbf{\textit{This work}}} \\
    \hline
       Element  & $\Delta$E$_{abs}$ & $\Delta$E$_{rel}$ (\%)  \\
    \hline
    \hline
    Se V & 0.4698 & 2.22 \\
        Se VI & 0.4466 & 1.87 \\
        Se VII & 1.8359 & 2.49 \\
        Se VIII & 1.0974 & 1.34 \\
        Se IX & / & / \\
        Se X & / & / \\
    \hline
    \end{tabular}
\end{table}

\reftab{tab: ErrorEstimation_Energy_SeV-X} shows that the accuracy of the Se V–VIII calculations is even higher than the Se I-IV ones. The precision of Se V and Se VII is within 2,5\%, while Se VI and Se VIII show deviations of less than 2\%.

As can be seen in \reffig{fig: energyBeyond_SeIV}, an exception has to be made for Se IX and X for which no \texttt{NIST} data are provided. In these two cases, the reliability of the results depends only on the quality of the level transitions calculated using the method described in \refsec{sec: GRASP2018_strategy}. 

In \reffig{fig: Full_Transition}, we report all transitions obtained from Se I to Se X with the corresponding classes.

\begin{figure}[htb!]
    \centering
    \includegraphics[width=1.0\linewidth]{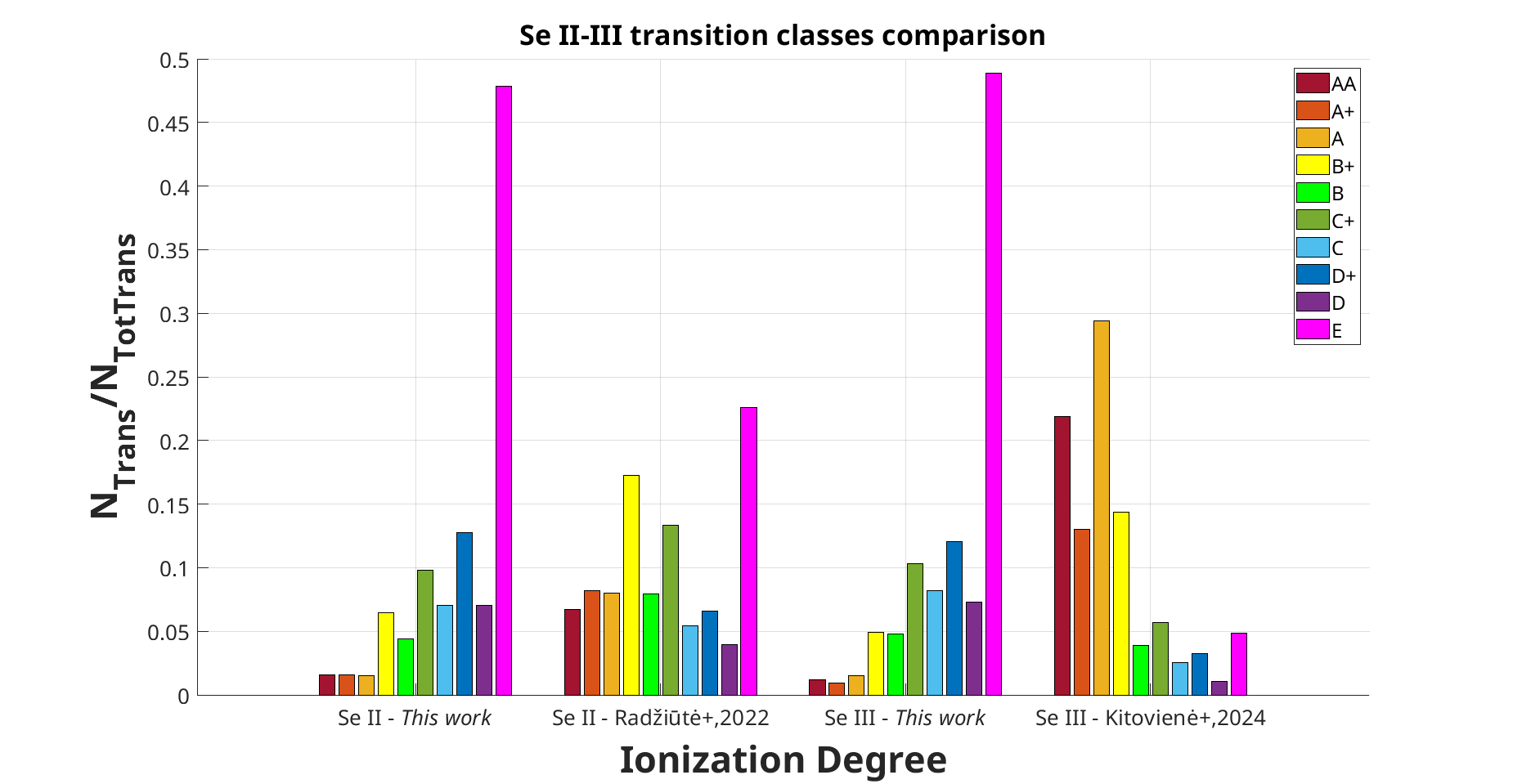}
    \caption{Se II-III transition classes comparison of \texttt{GRASP2018} calculation.}
    \label{fig: SeII-III_comparison}
\end{figure}

The classification proves that E-class transitions dominate at lower ionisation stages, while this trend reverses as we move toward higher stages. This indicates that the ASFs obtained for Se I–III are less accurate compared to those for Se IV–X. A similar conclusion can be reached by comparing our calculated transitions for Se II and Se III with the works of \citet{Radit2022} and \citet{Kitovien2024}, presented in \reffig{fig: SeII-III_comparison}.

\begin{figure*}[htb!]
    \centering
    \includegraphics[width=0.8\linewidth]{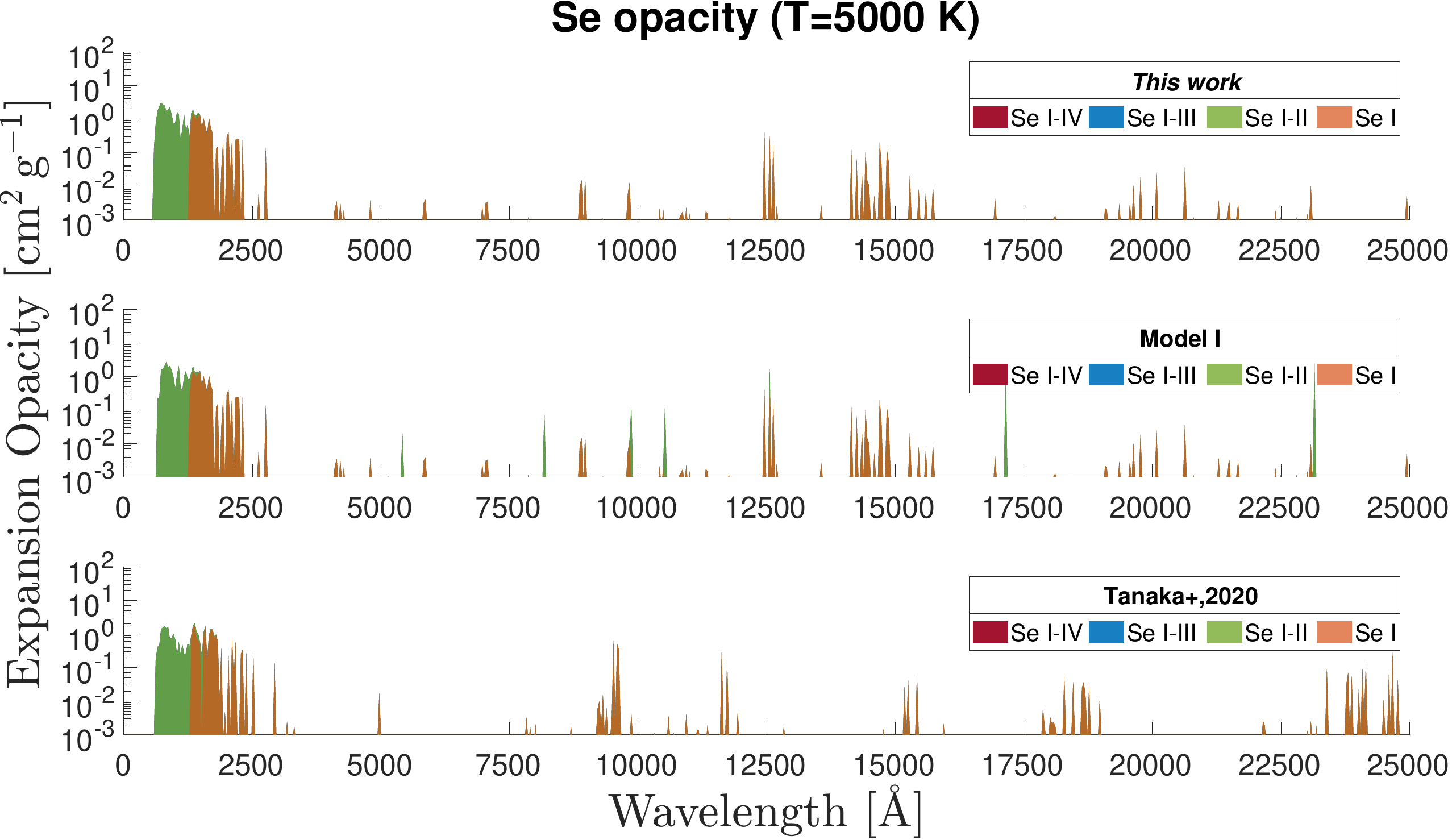}
    \caption{Comparison of Se I-IV expansion opacity at $\mathrm{T}=5\,000\,\mathrm{K}$ calculated with the three different set of atomic data. Each banner correspond to the expansion opacity calculated with the specific set of atomic data. (Top): the opacity is calculated considering Se I-IV data from model ``\textit{This work}''. (Middle): the opacity is calculated considering Se I-IV data from model ``Model I''. (Bottom): the opacity is calculated considering Se I-IV data from model ``Tanaka+,2020''. In plot legend, the notation ``I-II'', ``I-III'', and ``I-IV'' indicates that the opacity is calculated by including time-by-time an additional ionisation states.}
    \label{fig: SeI-IV_full_opacity_T5000}
\end{figure*}

However, in the early-stage scenario, the dominant ionisation stages are those beyond Se IV. Therefore, considering the high quality of these transitions (\reffig{fig: Full_Transition}), together with the high accuracy achieved on the energy levels (\reftab{tab: ErrorEstimation_Energy_SeV-X}), we provide a robust estimation for Se expansion opacity. Moreover, the Se I-X data presented and discussed in this section can be integrated into \texttt{MARTINI} platform to support future KN modelling.

\section{Expansion opacity}
\label{sec: ExpansionOpacity}

In the estimation of opacity, we included again the data of \citet{Radit2022}, \citet{Kitovien2024}, and \citet{Tanaka2020} to compare with our results of Se. In this context, in addition to our Se I-IV data, we build two other different models: the first is the previously mentioned ``Model I'', while the second is derived from the work of \citet{Tanaka2020}, with Se I-IV data collected from \citet{Lith-Jap}. In all models, the Se expansion opacity is estimated by assuming the LTE regime in the early-stage KN scenario. In detail, in \refeq{eq: ExpOpacity} we assume $\rho=10^{-13}\,\mathrm{g\,cm^{-3}}$, $t_{\mathrm{exp}}=1\,\mathrm{d}$, $\Delta \lambda$ = 34.50\,\AA\, at temperature values of $\mathrm{T}=5\,000\,\mathrm{K}, 10\,000\,\mathrm{K}$, while $\rho=3\times10^{-12}\,\mathrm{g\,cm^{-3}}$, $t_{\mathrm{exp}}=1\,\mathrm{d}$, $\Delta \lambda$ = 34.50\,\AA\, at $\mathrm{T}=20\,000\,\mathrm{K}$, and $100\,000\,\mathrm{K}$. Although the latter three values are significant temperatures for the early-stage KN, $\mathrm{T}=5\,000\,\mathrm{K}$ is also considered to perform an appropriate comparison of Se opacity with the results published in \citet{Tanaka2020}. \reffig{fig: SeI-IV_full_opacity_T5000} shows the Se I-IV opacity at $\mathrm{T}=5\,000\,\mathrm{K}$. In the plot, the three panels indicate the expansion opacity calculated for the three different models (i.e., ``\textit{This work}'', ``Tanaka+,2020'' and ``Model I''). 

In accordance with \reffig{fig: Se_ionisation_fraction}, all three models agree proving that Se II and Se I ionisation stages dominate at T = 5 000 K. In detail, all panels of \reffig{fig: SeI-IV_full_opacity_T5000} reproduce a maximum expansion opacity of $\kappa \sim 3.2\,\mathrm{cm^2\,g^{-1}}$, peaking in the wavelength range of 700 - 800\,\AA.
However, pushing to infrared wavelengths, the third model notably differs from the other two. Since all calculations were performed assuming the same thermodynamic conditions, the differences among the plots must be ascribed to the choice of Se I and Se II data used in the calculations. On the one side, the notable differences in \reffig{fig: SeI-IV_full_opacity_T5000} between ``\textit{This work}'' and ``Tanaka+,2020'' models are associated with the atomic datasets used for both Se I and Se II, whose energy levels accuracy (see \reftab{tab: ErrorEstimation_Energy}) reflects on the output opacity. On the other side, ``\textit{This work}'' and ``Model I'' share the same Se I data, and therefore, as can be evinced by the colours of \reffig{fig: SeI-IV_full_opacity_T5000}, the minor variations between the two must be attributed to the different Se II data considered, with the corresponding energy level accuracy and transition quality presented and discussed in \reftab{tab: ErrorEstimation_Energy} and Figs. \ref{fig: Full_Transition},\ref{fig: SeII-III_comparison}.

\begin{figure*}[htb!]
    \centering
    \includegraphics[width=0.8\linewidth]{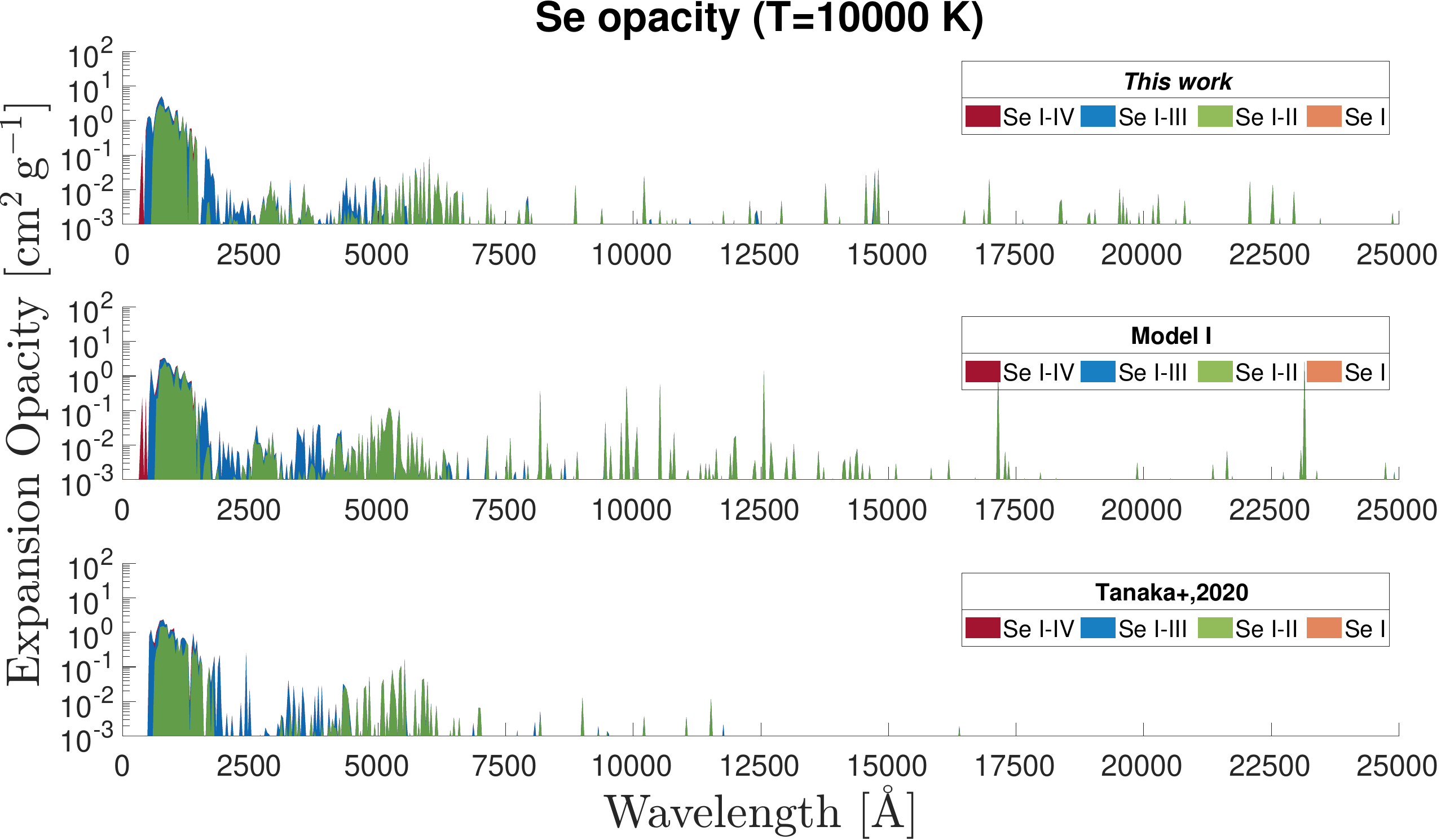}
    \caption{Comparison of Se I-IV expansion opacity at $\mathrm{T}=10\,000\,\mathrm{K}$ calculated with the three different set of atomic data. Each banner correspond to the expansion opacity calculated with the specific set of atomic data. (Top): the opacity is calculated considering Se I-IV data from model ``\textit{This work}''. (Middle): the opacity is calculated considering Se I-IV data from model ``Model I''. (Bottom): the opacity is calculated considering Se I-IV data from model ``Tanaka+,2020''. In plot legend, the notation ``I-II'', ``I-III'', and ``I-IV'' indicates that the opacity is calculated by including time-by-time an additional ionisation states.}
    \label{fig: SeI-IV_full_opacity_T10000}
\end{figure*}

We extend the analysis to higher temperatures of interest for the early KN scenario. \reffig{fig: SeI-IV_full_opacity_T10000} shows the expansion opacity estimation of the three models at $\mathrm{T}=10\,000\,\mathrm{K}$. The ionisation states mainly contributing to the opacity are Se II and Se III, in agreement with \reffig{fig: Se_ionisation_fraction}. Moreover, all three models exhibit a high opacity contribution in the range of 600-2\,000\,\AA\, with a peak around $1\,500\,$\AA\, of $\kappa \sim  10\,\mathrm{cm^2\,g^{-1}}$ value. A secondary opacity peak of $\kappa \sim 0.02\,\mathrm{cm^2\,g^{-1}}$ can also be observed around 5\,500-6\,000\,\AA, while beyond $\sim 7\,000\,$\AA, the three plots differ in behaviour. This is a consequence of Se II and Se III data included in the models. Their deviation from the \texttt{NIST ASD} presented in \reftab{tab: ErrorEstimation_Energy} affect the opacity estimation, producing different trends among the plots. The same conclusions can be drawn between ``\textit{This work}'' and ``Model I'' comparison: even if the plots look similar, the different accuracy on the transition quality evaluation reported in \reffig{fig: SeII-III_comparison} reflects on the opacity. This can be seen by the presence of green peaks induce by Se II in mid-panel, not observable on the other. Finally, unlike bottom plot, top- and mid-panels report a small contribution of Se IV at T = $10\,000\,\mathrm{K}$. Again, the Se IV implemented in ``\textit{This work}'' and ``Model I'' is exactly the same and results to be more accurate than the one of ``Tanaka+,2020'' (see \reftab{tab: ErrorEstimation_Energy}).

\begin{figure*}
    \centering
    \begin{minipage}{0.47\textwidth}
    \centering
    \phantomsection
    \hypertarget{fig: Se20000K}{}
        \includegraphics[width=\linewidth]{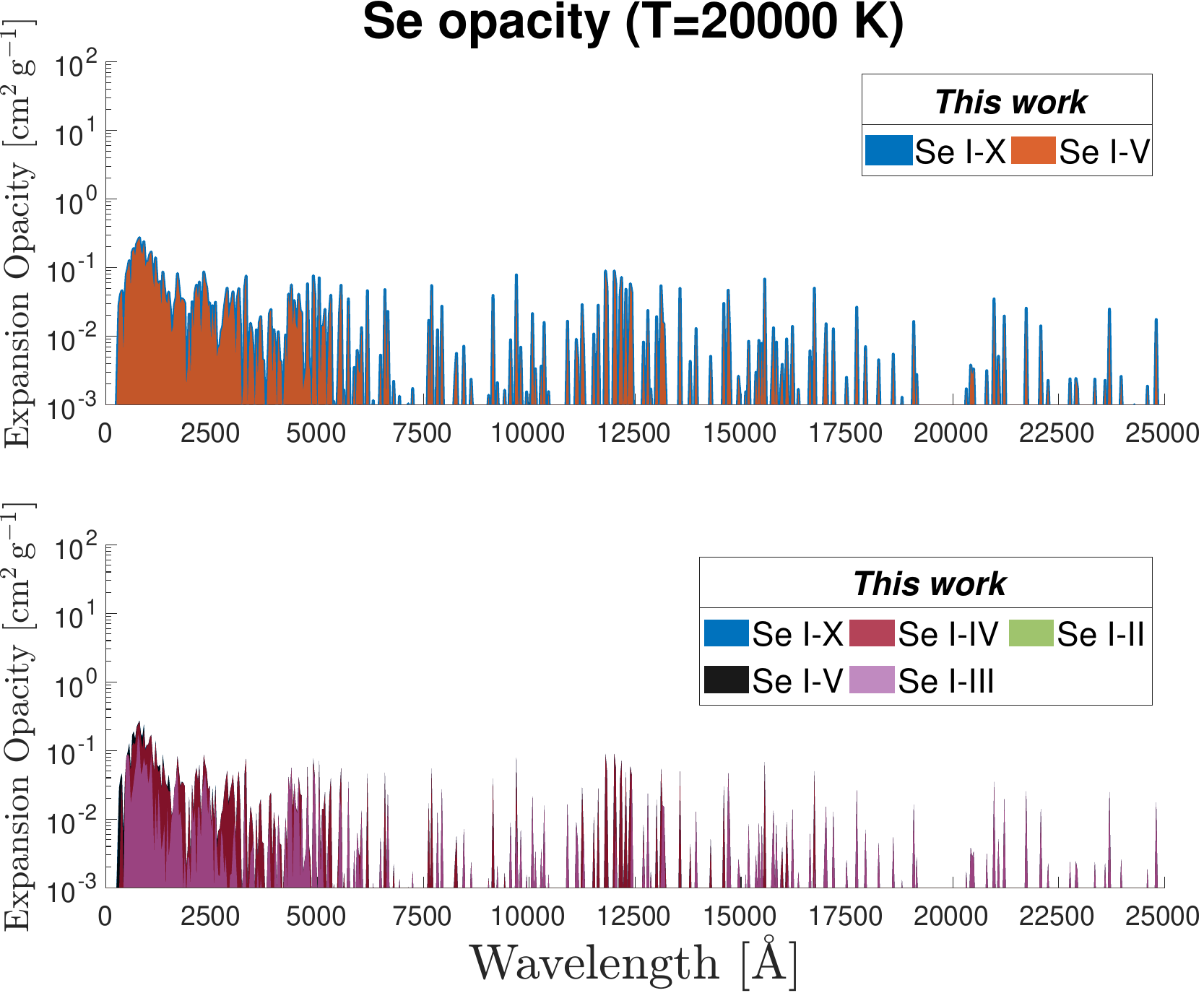}
        \vspace{0.5ex}
        \small \textbf{(a)} 
    \end{minipage}
    \begin{minipage}{0.47\textwidth}
    \centering
    \phantomsection
    \hypertarget{fig: Se20000K_zoom}{}
        \includegraphics[width=\linewidth]{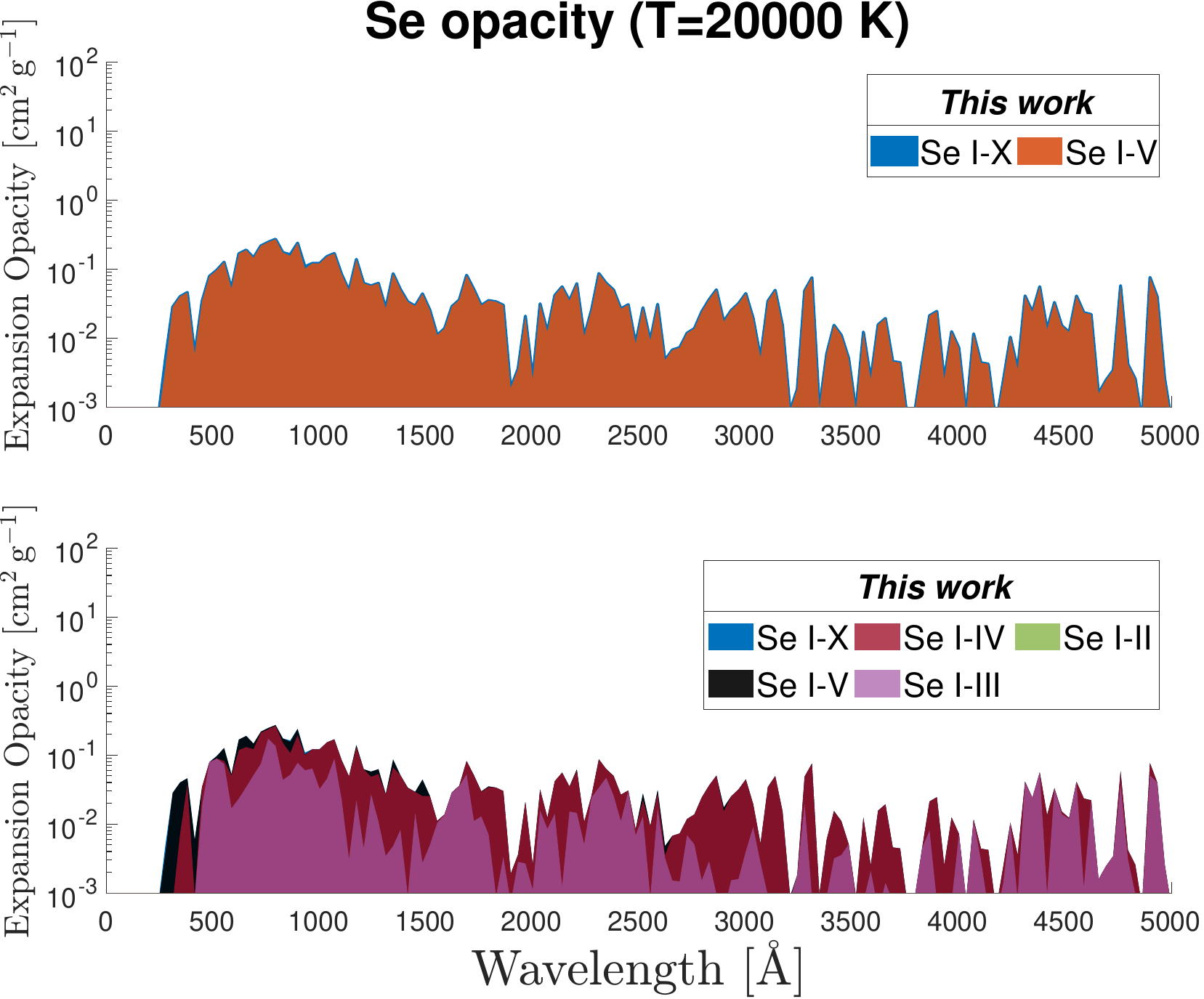}
        \vspace{0.5ex}
        \small \textbf{(b)} 
    \end{minipage}
    \caption{Comparison of Se I-X expansion opacity at $\mathrm{T}=20\,000\,\mathrm{K}$ calculated with Se I-X data presented in this work. In plot legend, the notation ``I-X'' indicates that the opacity is calculated by including all the ionisation states from ``I'' to ``X''. On top panel of (a), it is shown a comparison between the Se I-X and Se I-V ionisation states, while on bottom panel it is presented the different ionisation state contribution from Se I to Se X. Plot (b) shows the same analysis with a zoom-in-view in the wavelength range 0-5\,000\,\AA. From both figures, it can be seen that only the ionisation degrees between Se III and Se V matters on the opacity calculation at T = 20\,000\,K.}
    \label{fig: SeT20000_opacity}
\end{figure*}

As shown in \reffig{fig: Se_ionisation_fraction}, in early-stage KN scenario temperatures can reach higher values beyond $\mathrm{T}=10\,000\,\mathrm{K}$. For this reason, the second temperature investigated is $\mathrm{T}=20\,000\,\mathrm{K}$. In this case, \reffig{fig: Se_ionisation_fraction} shows that among the dominant ionisation states there is also Se V. However, no additional Se V data are available from the literature. Therefore, in the analysis we use only the Se I-X data presented in this work.

\reffig{fig: SeT20000_opacity}~\hyperlink{fig: Se20000K}{(a)} shows the expansion opacity behaviour at $\mathrm{T}=20\,000\,\mathrm{K}$. The top panel shows the comparison between the expansion opacity calculated for all Se I-X data, compared to the one calculated with Se I-V only. These results are in agreement with \reffig{fig: Se_ionisation_fraction}, demonstrating that, at $\mathrm{T}=20\,000\,\mathrm{K}$, all contributions beyond Se V are negligible. In the bottom panel, it is shown how the different ionisation stages from Se I to Se V impact in the opacity. Specifically, \reffig{fig: SeT20000_opacity}~\hyperlink{fig: Se20000K_zoom}{(b)} represents a zoom-in-view of \reffig{fig: SeT20000_opacity}~\hyperlink{fig: Se20000K}{(a)} in the wavelength range 0-5\,000\,\AA. From the plot, it can be seen that below Se III there are no Se ionisation degrees relevant at $\mathrm{T}=20\,000\,\mathrm{K}$. Moreover, with the increase of the ionisation states, the opacity is shifting from the visible towards the near-ultraviolet region.

Things become even more evident when we push to $\mathrm{T}=100\,000\,\mathrm{K}$. In \reffig{fig: SeI-IV_full_opacity_T100000}, we show a comparison of the Se I-X at $\mathrm{T}=100\,000\,\mathrm{K}$. It can be seen that, at this high temperature, there is no opacity contribution below Se VI. Only the ionisation stages beyond Se VI have an impact on producing evident traces in the observed opacity plot. At $\mathrm{T}=100\,000\,\mathrm{K}$, the opacity of Se VIII-X can be detected by two peaks of $\kappa \sim 1-1.4\,\mathrm{cm^2\,g^{-1}}$, located in the ultraviolet region within 100 - $2\,000\,$\AA. 

Modest variations in the thermodynamic parameters within the range considered slightly affect the absolute opacity, but its overall behaviour remains unchanged. Therefore, the opacity results presented are robust within the temperature–density range investigated under LTE conditions.

\section{\texttt{POSSIS} spectral analysis}
\label{sec: SpectraAnalysis}

The last part of the work concerns the analysis of KN spectra through the \texttt{POSSIS} code. 
Since \texttt{POSSIS} requires a pre-computed table of opacity as input (see \refsec{sec: theory_POSSIS}), for the analysis, we calculate three different grids of opacity values using the Se atomic data of the three models (i.e., \textit{This work}, \citet{Tanaka2020} data and ``Model I'') presented in \refsec{sec: SeResults}. In the opacity estimation, we consider density values ranging from -19.5 to -4.5 $\mathrm{g\,cm^{-3}}$ on the logarithmic scale at steps of 0.5 $\mathrm{g\,cm^{-3}}$ and temperatures within $1\,000$ and $51\,000\,$ K with an increase of 500 K at each density iteration.

In the model, we assume spherically symmetric ejecta, with a mass of $M_{\mathrm{ej}}$ = 10$^{-2}\,$M$_{\odot}$ which is comparable to the estimated mass of AT2017gfo, an averaged velocity of $v_{\mathrm{ej}}=0.2\,\mathrm{c}$ and a fixed electron fraction of $Y_{\mathrm{e}}$ = 0.35, representative of the ejected matter responsible for the early KN emission. \texttt{POSSIS} simulations are run for a total number of photon packets $N_{\mathrm{ph}}$ = 10$^7$. We focus on three different time epochs after the BNS merger: $0.5\,\mathrm{d}$, $1\,\mathrm{d}$, and $1.43\,\mathrm{d}$. In all time epochs, we fixed $Y_\mathrm{e}$=0.35 to select the appropriate heating-rate, and we investigated how the precomputed Se opacity influences the KN spectrum. The analysis is carried out by comparing three spectra generated by the Se opacity results of the three different models presented in the previous section. Moreover, in every epoch two scenarios are considered. The first scenario resembles a KN in which the opacity contribution comes from an ejecta consisting of 100\% Se. In the second scenario, the Se opacity contributes only partially to the total opacity, since it is taken to be a fraction of 10\% of the total mass (see \reffig{fig: Se_ionisation_fraction}). The remaining 90\% contributes with an adopted gray opacity of $0.5\,\mathrm{cm^2\,g^{-1}}$, typical for lanthanide-poor compositions \citep{Villar2017}. 

To assess whether Se could produce observable spectral features in kilonova ejecta, we perform a simple test by combining the precomputed pure-Se expansion opacity $\kappa_\mathrm{Se}^\mathrm{pure}$ with a grey mixture opacity $\kappa_\mathrm{grey}^\mathrm{mix}$=0.5 $\,\mathrm{cm^2\,g^{-1}}$. Specifically, in the case where Se constitutes the 100\% of KN ejecta mass, the selenium abundance is assumed as $X_\mathrm{Se}$=1 and therefore the whole KN opacity grid is Se opacity, constructed as
\begin{equation}
    \kappa_\mathrm{KN}^{100\%} = X_\mathrm{Se}\,\kappa_\mathrm{Se}^\mathrm{pure}.
    \label{eq: k_Sepure}
\end{equation}
In the case of Se at 10\%, the Se abundance becomes $X_\mathrm{Se}$=0.1, consistent with expectations for high-Ye ejecta (see \reffig{fig: Ye_trajectories}). In this scenario, \refeq{eq: k_Sepure} is modified as
\begin{equation}
    \kappa_\mathrm{KN}^{10\%} = (1 - X_\mathrm{Se})\,\kappa_\mathrm{grey}^\mathrm{mix}+X_\mathrm{Se}\,\kappa_\mathrm{Se}^\mathrm{pure} \ .
    \label{eq: k_trial}
\end{equation}
From \refeq{eq: k_trial}, whenever $\kappa_\mathrm{KN}^{10\%}$ < $\kappa_\mathrm{grey}^\mathrm{mix}$, the perturbative effects caused by Se opacity are not detectable in the resulting spectra. Only in the case of $\kappa_\mathrm{KN}^{10\%}$ > $\kappa_\mathrm{grey}^\mathrm{mix}$, possible Se-induce feature could be observed. A caveat must be noted: the heating rates should be distinguished and treated separately for each scenario; however, since here we are interested in the impact of different opacities on the final spectra, we adopted the same heating rates in both cases.

\begin{figure*}
    \centering
    \begin{minipage}{0.45\textwidth}
    \centering
    \phantomsection
    \hypertarget{fig: KN_t0.5_100}{}
        \includegraphics[width=\linewidth]{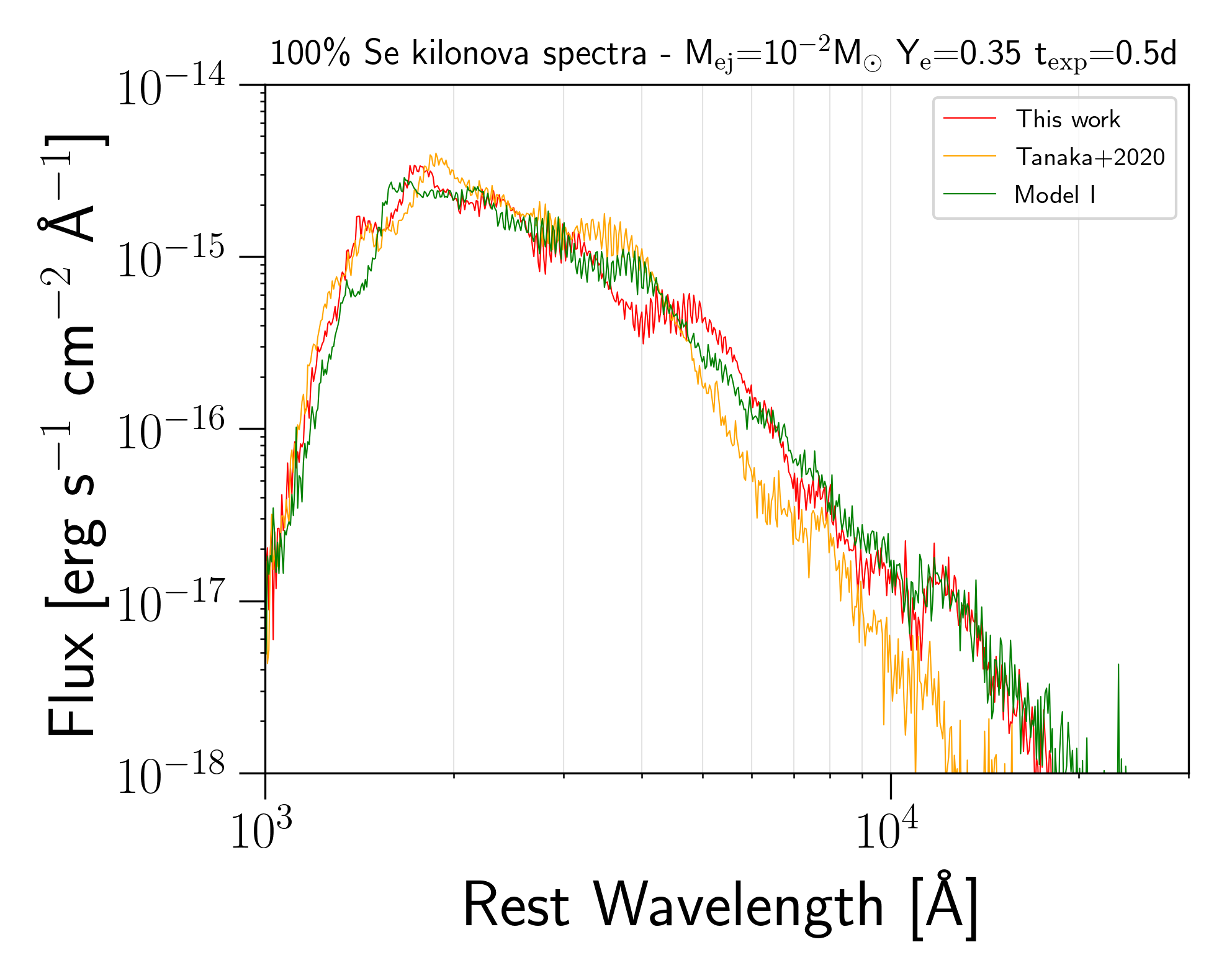}
        \vspace{0.5ex}
        \small \textbf{(a)} KN spectra at 40 Mpc of distance with 100\% Se composition at $t_{\mathrm{exp}}=0.5\,\mathrm{d}$.
    \end{minipage}
    \begin{minipage}{0.45\textwidth}
    \centering
    \phantomsection
    \hypertarget{fig: KN_t0.5_10}{}
        \includegraphics[width=\linewidth]{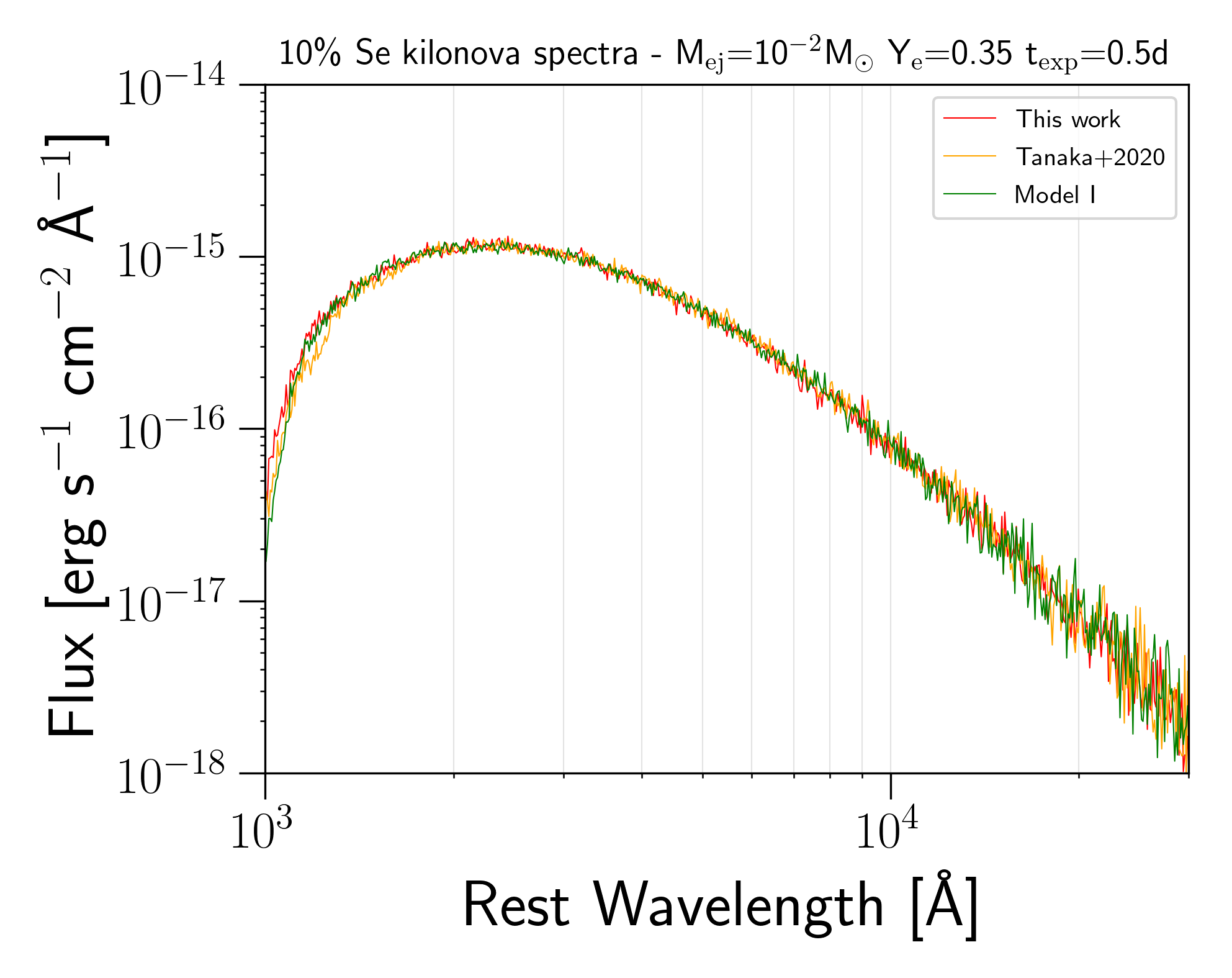}
        \vspace{0.5ex}
        \small \textbf{(b)} KN spectra at 40 Mpc of distance with 10\% Se composition at $t_{\mathrm{exp}}=0.5\,\mathrm{d}$.
    \end{minipage}
    \begin{minipage}{0.45\textwidth}
    \centering
    \phantomsection
    \hypertarget{fig: KN_t1_100}{}
        \includegraphics[width=\linewidth]{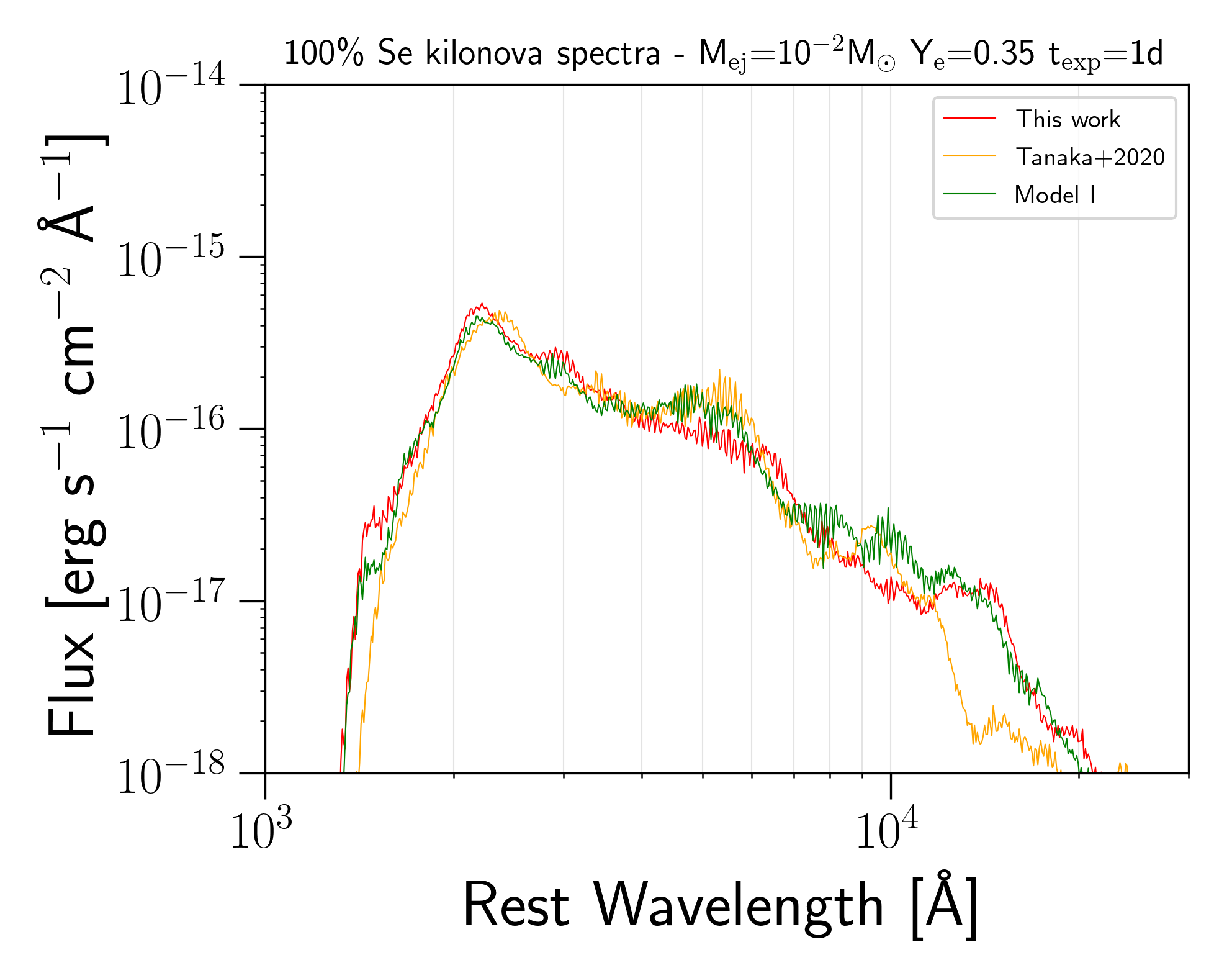}
        \vspace{0.5ex}
        \small \textbf{(c)} KN spectra at 40 Mpc of distance with 100\% Se composition at $t_{\mathrm{exp}}=1\,\mathrm{d}$.
    \end{minipage}
    \begin{minipage}{0.45\textwidth}
    \centering
    \phantomsection
    \hypertarget{fig: KN_t1_10}{}
        \includegraphics[width=\linewidth]{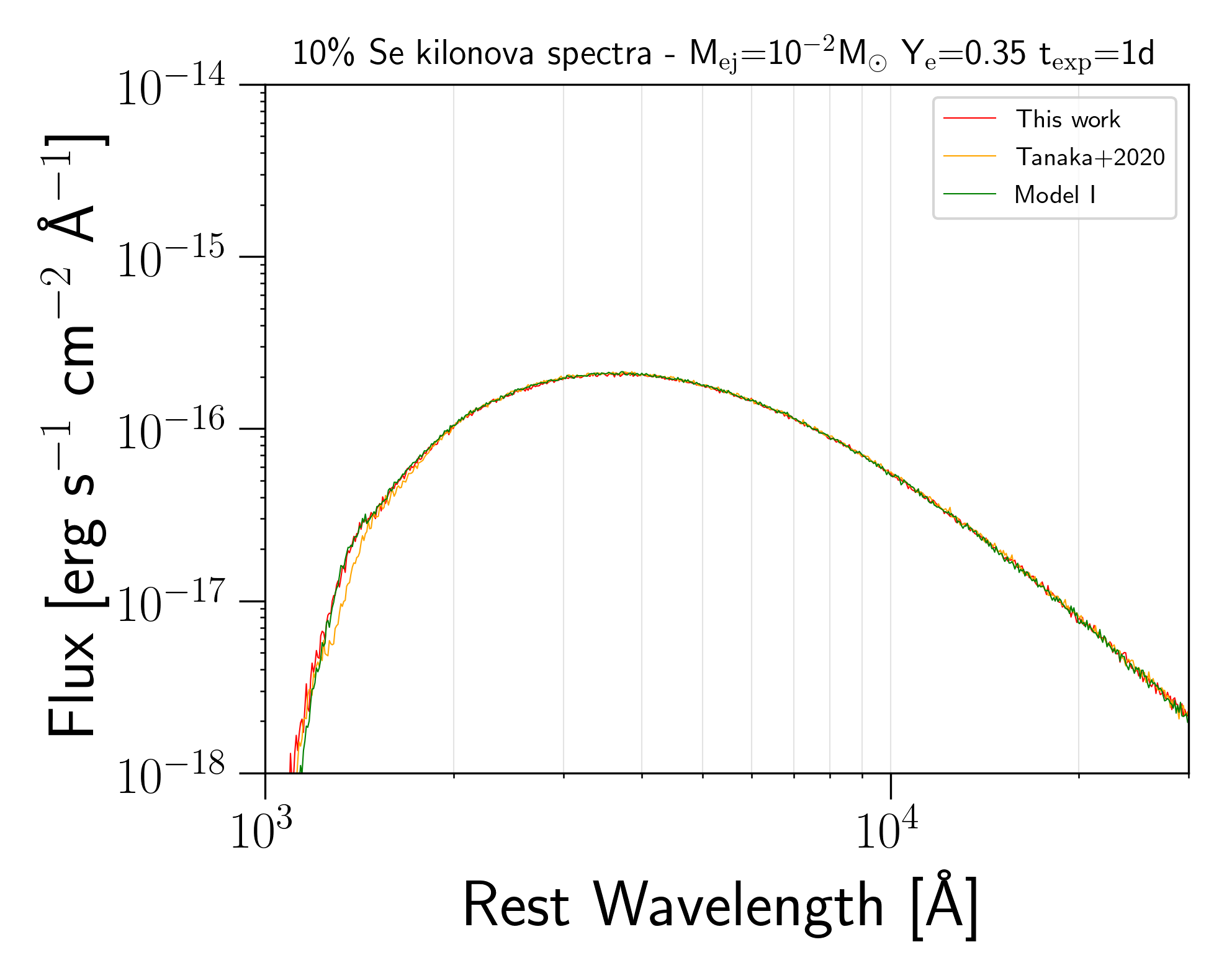}
        \vspace{0.5ex}
        \small \textbf{(d)} KN spectra at 40 Mpc of distance with 10\% Se composition at $t_{\mathrm{exp}}=1\,\mathrm{d}$.
    \end{minipage}
    \begin{minipage}{0.45\textwidth}
    \centering
    \phantomsection
    \hypertarget{fig: KN_t1.43_100}{}
        \includegraphics[width=\linewidth]{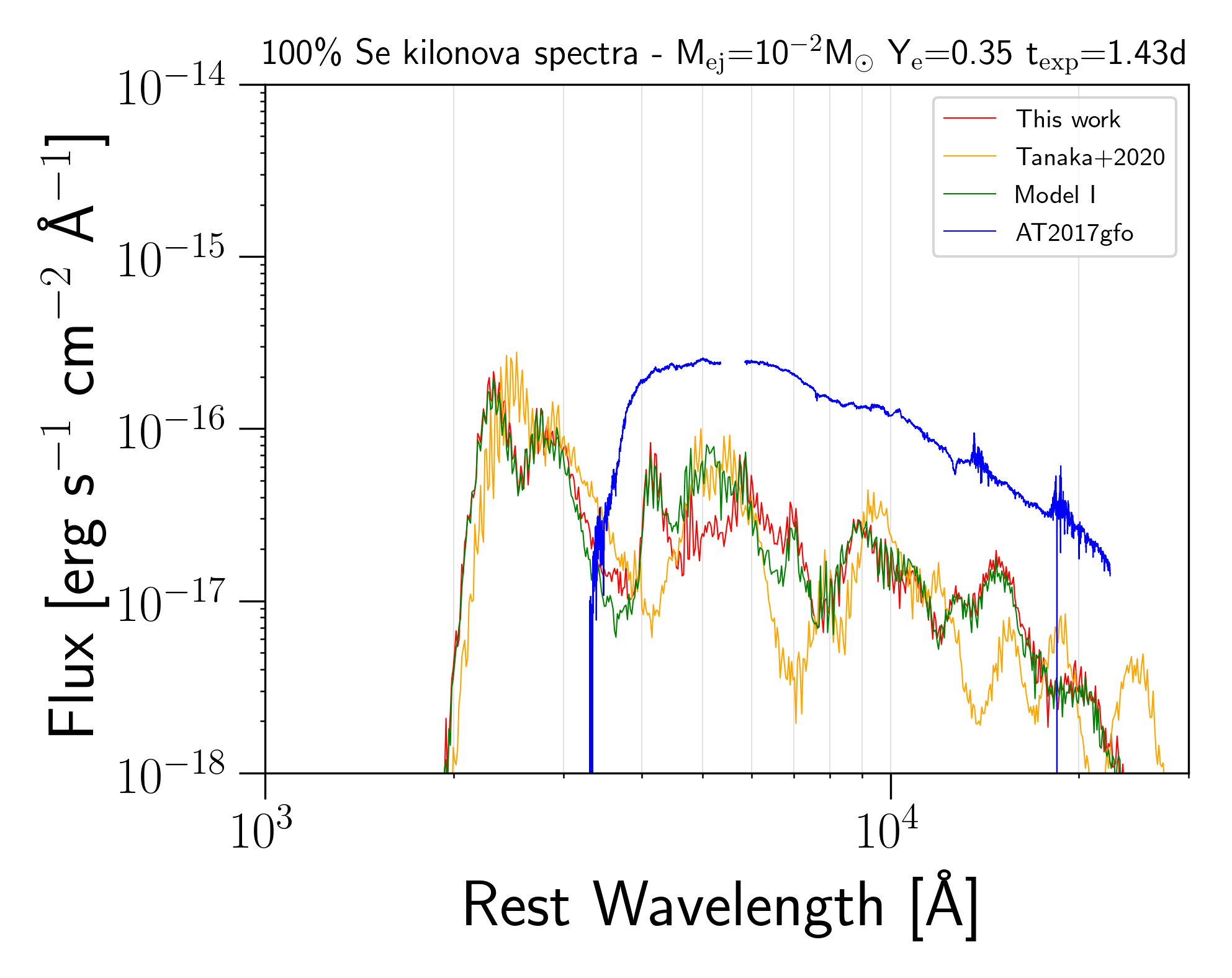}
        \vspace{0.5ex}
        \small \textbf{(e)} KN spectra at 40 Mpc of distance with 100\% Se composition at $t_{\mathrm{exp}}=1.43\,\mathrm{d}$.
    \end{minipage}
    \begin{minipage}{0.45\textwidth}
    \centering
    \phantomsection
    \hypertarget{fig: KN_t1.43_10}{}
        \includegraphics[width=\linewidth]{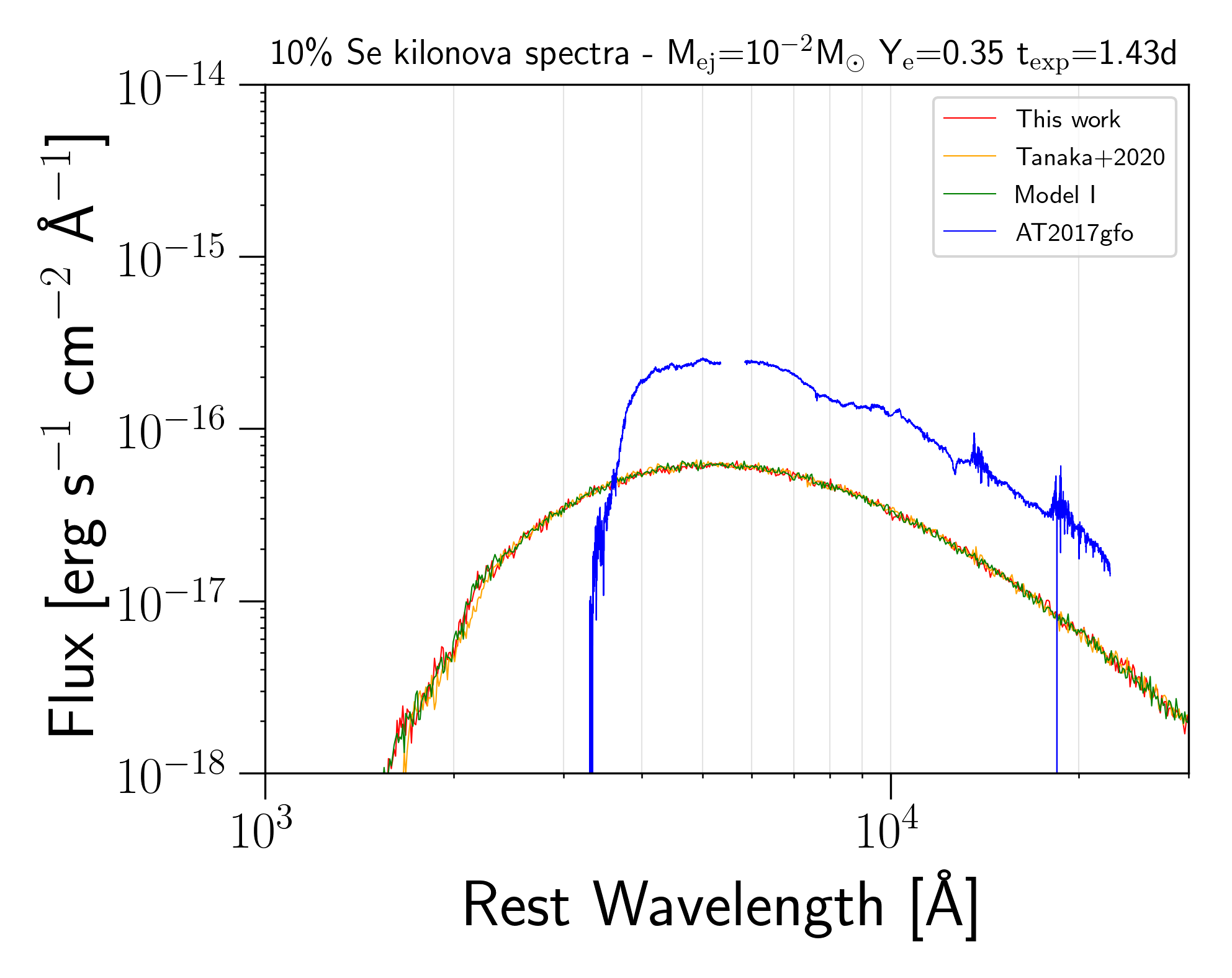}
        \vspace{0.5ex}
        \small \textbf{(f)} KN spectra at 40 Mpc of distance with 10\% Se composition at $t_{\mathrm{exp}}=1.43\,\mathrm{d}$.
    \end{minipage}
    \caption{KN spectra comparison between the three different models, generated through the \texttt{POSSIS} code. Each row represents a different time epoch: $0.5\,\mathrm{d},\,1\,\mathrm{d},\,1.43\,\mathrm{d}$, respectively. The column identify different KN ejecta composition: on the left column, we compared the spectra generated by an opacity contribution of 100\% of Se. The right column was calculated by considering Se at 10\%, and a gray opacity $\kappa_{\mathrm{gray}}^\mathrm{mix}=0.5\,\mathrm{cm^2\,g^{-1}}$ for the remaining 90\% of the composition. In the last epoch, we also included the AT2017gfo spectrum for a consistent comparison in the analysis. }
    \label{fig: POSSIS_Spectra_Full}
\end{figure*}

Figure \ref{fig: POSSIS_Spectra_Full} shows the comparison of the two KN scenarios at three different epochs: $t_{\mathrm{exp}}=0.5, 1.0, 1.43\,\mathrm{d}$. The third epoch is chosen to facilitate comparison with the first X-shooter data for AT2017gfo \citep{pian2017,Smartt2017,Tanvir2017}. A clear trend emerges: the spectra for the model with 100\% Se peak at shorter wavelengths than those of the 10\% model, with the difference growing in time. This effect is driven by the different opacities between the two models. In the 10\% Se model the inclusion of an additional gray opacity of 0.5\,cm$^2$\,g$^{-1}$ increases the overall opacity relative to the 100\% Se case. As a result, the thermalisation/photospheric surface in the 100\% Se model lies deeper inside the ejecta in regions that are hotter. This is illustrated in \reffig{fig: Tph} at 1\,d after the merger. The surface where the thermalisation depth $\tau_{\mathrm{eff}}=\sqrt{\tau_{abs}\,(\tau_{abs}+\tau_{sc})}\sim 1$ \citep{Rybicki1979} is located at velocities of $\sim0.05$c in the 100\% Se model and at $\sim0.12$c in the 10\% Se model. The corresponding temperatures are $\sim 11\,200$\,K for the 100\% Se model and $7\,500$\,K for the 10\% Se model. Black-body spectra at these temperatures peak at $\sim 2\,700\,\AA$ and $\sim 4\,000\,\AA$, in good agreement with the modelled spectra at 1 day (see \reffig{fig: POSSIS_Spectra_Full}~\hyperlink{fig: KN_t1_100}{(c)} and \reffig{fig: POSSIS_Spectra_Full}~\hyperlink{fig: KN_t1_10}{(d)}).

We now look more closely at the spectra at each epoch and focus on the differences between the three models with different Se opacities. \reffig{fig: POSSIS_Spectra_Full}~\hyperlink{fig: KN_t0.5_100}{(a)} and \reffig{fig: POSSIS_Spectra_Full}~\hyperlink{fig: KN_t0.5_10}{(b)} show the comparison of the two KN scenario at $t_{\mathrm{exp}}=0.5\,\mathrm{d}$. Focusing on the left-hand side plot, we find that the spectra shape of the three models is similar. On the one hand, the curves show a tail beyond $\sim5\,000\,$\AA, with the highest value about ${4}\times10^{-15}\,\mathrm{erg\, s^{-1}\,cm^{-2}\,}$\AA${^{-1}}$ near $2\,000\,$\AA. On the other hand, all the three models predict a Se spectral feature in the range of $3\,000 - 4\,000\,$\AA. Further differences are also notable among these spectra. The ``peaks'' do not perfectly overlap, and the absolute intensities vary from spectrum to spectrum. Since we assumed the same modelling criteria, these variations are a consequence of the different opacity inputs adopted in each model. Changing the opacities affects the propagation of photon packets, and consequently the radiation field (Eq.~\ref{eq:MCest}), which in turn influences the temperature (Eq.~\ref{eq:SB}) and the opacities themselves. An example of this effect is shown in Fig.~\ref{fig: Tph}. These differences in the opacities and the uncertainties in the atomic calculations are reflected in the resulting spectra.

If we now focus on the right-hand plot (\reffig{fig: POSSIS_Spectra_Full}~\hyperlink{fig: KN_t0.5_10}{(b)}), it can be seen that in a more ``realistic'' scenario in which Se opacity contributes only on a 10\% of the whole KN mass, the shape of the spectra results in a more regular behaviour with no Se-induced features observable. This is a consequence of \refeq{eq: k_trial}. In the previous section, we have seen that Se higher opacity is always located in the near-ultraviolet/visibile range, while it becomes almost negligible at large wavelengths, in the near-infrared region. This behaviour is reflected in the spectra. In the wavelength range beyond 2\,000\,\AA, $\kappa_{\mathrm{Se}}^\mathrm{pure}$ reaches small values such that $\kappa_{\mathrm{KN}}^\mathrm{10\%}$ opacity results to be smaller than $\kappa_{\mathrm{gray}}^\mathrm{mix}$. Therefore, the grey opacity overwhelm the Se contribution, with no detectable features on the spectra. On the contrary, the slight differences between the models that can be observed at low wavelengths are produced by $\kappa_{\mathrm{Se}}^\mathrm{pure}$ that contributes to exceed the grey opacity. In this case $\kappa_{\mathrm{KN}}^\mathrm{10\%}$ becomes greater than $\kappa_{\mathrm{grey}}^\mathrm{mix}$ with the consequence of slight spectra distortion from the continuum shape.

In the epoch $t_{\mathrm{exp}}=1\,\mathrm{d}$, similar spectra characteristics can be observed. The results are shown in \reffig{fig: POSSIS_Spectra_Full}~\hyperlink{fig: KN_t1_100}{(c)} and \reffig{fig: POSSIS_Spectra_Full}~\hyperlink{fig: KN_t1_10}{(d)}. In general, the left-hand plot shows good agreement between our model (red) and ``Model I'' (green) if compared to the \citet{Tanaka2020} model (orange). All three predict a highest intensity value of 6$\times$10$^{-16}\mathrm{erg\, s^{-1}\,cm^{-2}\,}$\AA${^{-1}}$ at $\sim2\,200\,$\AA, but differ at $\sim10\,000\,$\AA. Moving to the second scenario, on \reffig{fig: POSSIS_Spectra_Full}~\hyperlink{fig: KN_t1_10}{(d)} no evident discrepancies between the models can be detected, beside the small variations between the plots at low wavelengths, in agreement with the previous analysis.

In the late epoch of $t_{\mathrm{exp}}=1.43\,\mathrm{d}$, significant differences between the models can be observed. We report them in \reffig{fig: POSSIS_Spectra_Full}~\hyperlink{fig: KN_t1.43_100}{(e)} and \reffig{fig: POSSIS_Spectra_Full}~\hyperlink{fig: KN_t1.43_10}{(f)}, together with the uniform reduction and calibration of the X-shooter data for AT2017gfo at the same epoch \citep{pian2017,Smartt2017,Tanvir2017}.

In the left plot, the shapes of the red and green models differ greatly from those of the orange model. In particular, there are absorption features in the “Tanaka+,2020” model that are not there in the other two. In detail, the spectrum calculated with our Se opacity (red) agrees better with the spectrum of ``Model I'' (green) generated with the most accurate Se atomic data available. Both spectra show less absorption behaviour at $4\,000\,$\AA\, and $7\,000\,$\AA, compared to the model of \citet{Tanaka2020} which predicts a much stronger absorption feature. At $5\,000\,$\AA, ``Model I'' and \citet{Tanaka2020} both provide an intensity ``peak'' of 10$^{-16}\mathrm{erg\, s^{-1}\,cm^{-2}\,}$\AA${^{-1}}$, while, in our model, it is shown at a shifted wavelength position of $6\,000\,$\AA. These different positions are a direct consequence of the estimate opacity and the accuracy of the atomic calculations.

Finally, a gap in intensity can be noticed between the predicted spectra of the three models and the observed spectrum of AT2017gfo. This difference can be attributed to the over-simplified model of one spherical symmetric component of 10$^{-2}\mathrm{M}_{\odot}$ assumed in the analysis. In a more realistic scenario, at the epoch of $t_{\mathrm{exp}}=1.43\,\mathrm{d}$ multiple components of the ejecta with different Y$_{\mathrm{e}}$ and a greater total mass of $5\times10^{-2}\mathrm{M}_{\odot}$ are likely needed to explain the AT2017gfo spectra \citep{Villar2017,Perego2017}. In addition to the difference in intensity, the modelled spectra are typically bluer than the one of AT2017gfo, especially for the case of 100\% Se. This effect is driven by the opacities in the 100\% Se model being much lower than those expected in KN ejecta and therefore producing spectra that are bluer than those observed (cf. with the photospheric temperature of $\sim5000$\,K estimated for AT2017gfo in \citealt{Gillanders2022}). The temperature of the 10\% Se model is more similar to the observed one. However, the model does not show a strong flux suppression below $\sim4000$\,\AA{} as in AT2017gfo since the adopted gray opacity is only a crude approximation and does not capture the strong line blanketing expected at these phases in these blue wavelength regions (see e.g. fig. 11 of \citealt{Tanaka2020} where the opacity at 1 day and at these wavelengths can reach $\sim10-50$ cm$^2$\,g$^{-1}$ even for $Y_e=0.35$). 

In the plot of \reffig{fig: POSSIS_Spectra_Full}~\hyperlink{fig: KN_t1.43_10}{(f)}, it can be seen that all three models perfectly overlap and therefore no possible features of Se can be detected.

\section{Selenium Atomic Opacities in the \texttt{MARTINI} Platform}

\begin{figure*}
    \centering
    \begin{minipage}{0.42\textwidth}
    \centering
    \phantomsection
    \hypertarget{fig: MainMenu}{}
        \includegraphics[width=\linewidth]{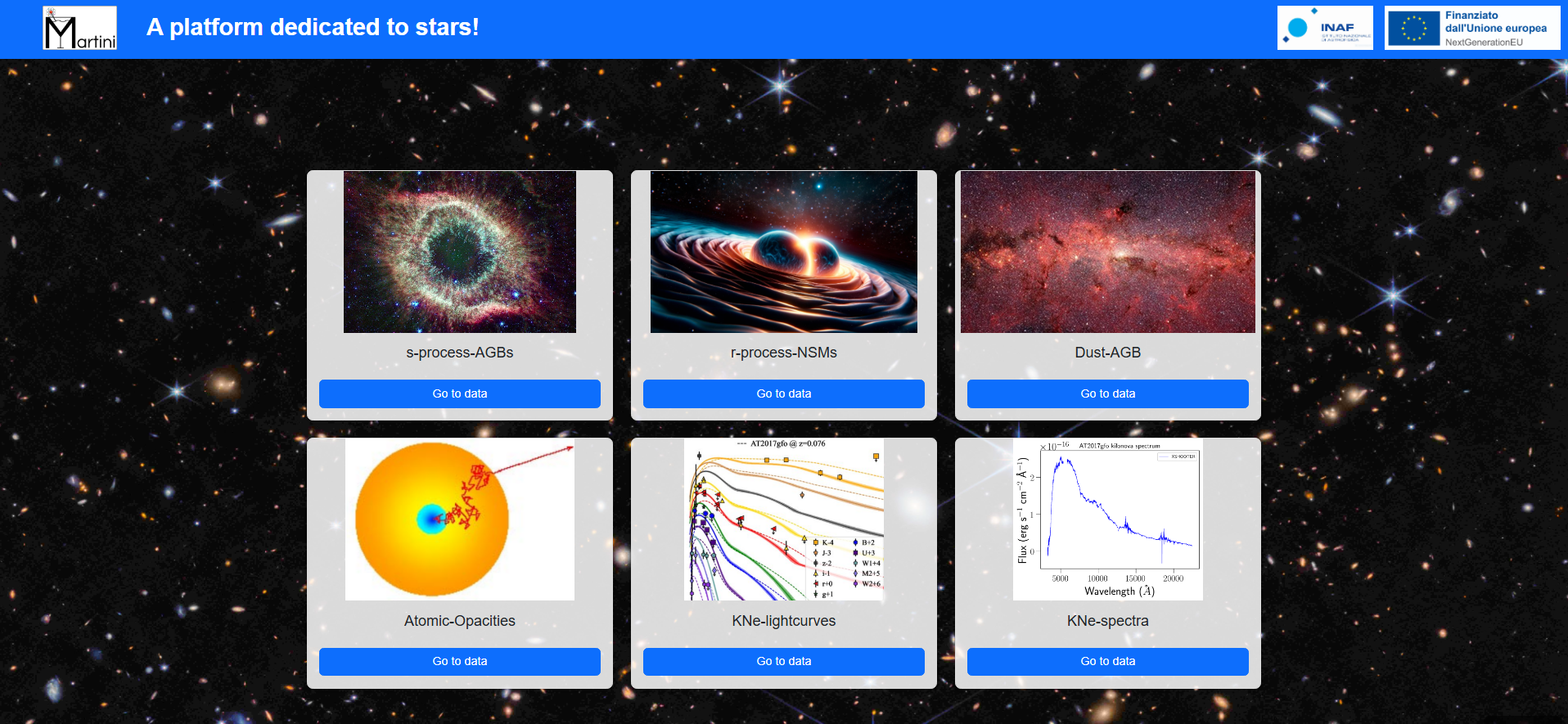}
        \vspace{0.5ex}
        \small \textbf{(a)} Homepage of the \texttt{MARTINI} platform.
    \end{minipage}
    \begin{minipage}{0.4\textwidth}
    \centering
    \phantomsection
    \hypertarget{fig: PeriodicTable}{}
        \includegraphics[width=\linewidth]{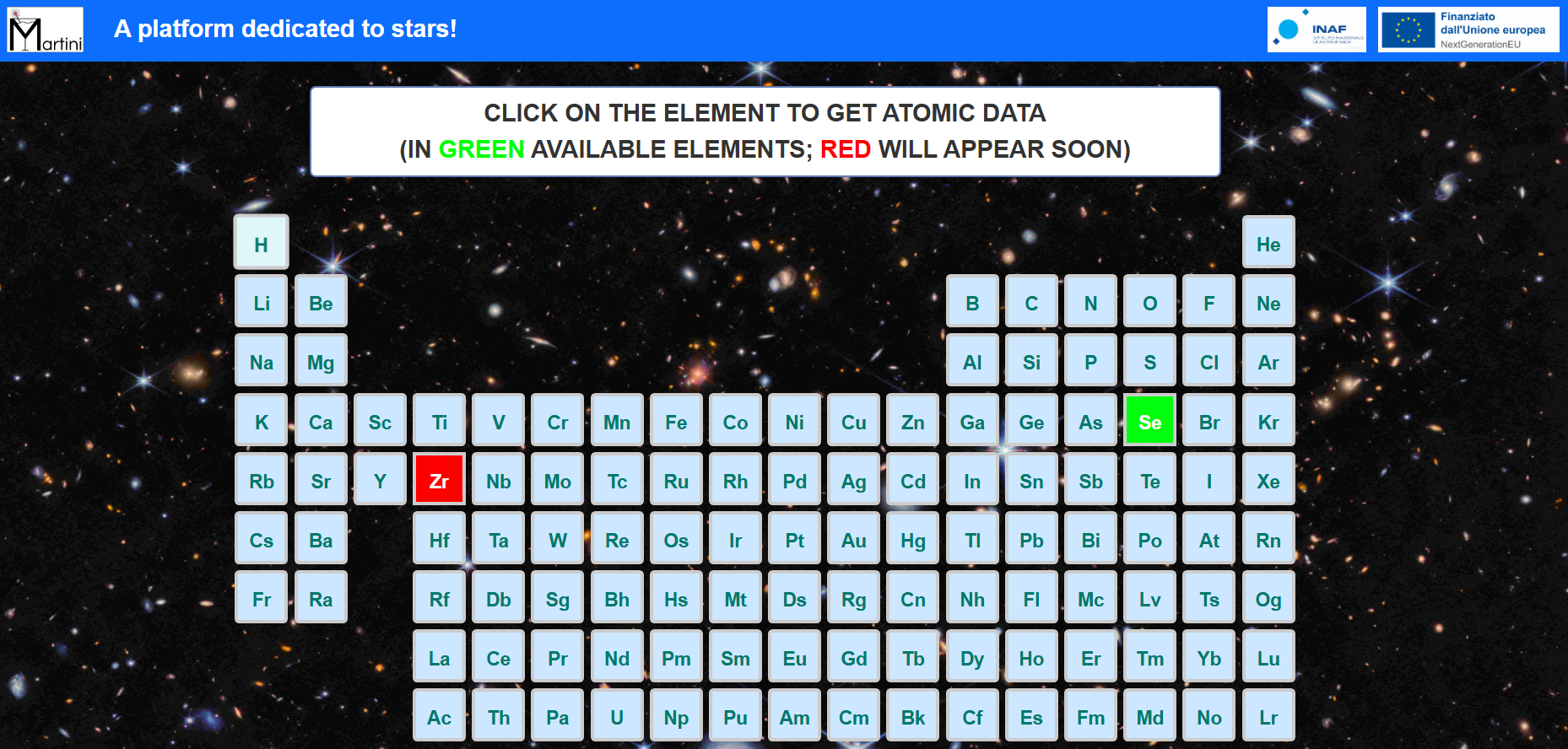}
        \vspace{0.5ex}
        \small \textbf{(b)} Periodic table for the element selection.
    \end{minipage}
    \begin{minipage}{0.8\textwidth}
    \centering
    \phantomsection
    \hypertarget{fig: AtomicDetail}{}
        \includegraphics[width=\linewidth]{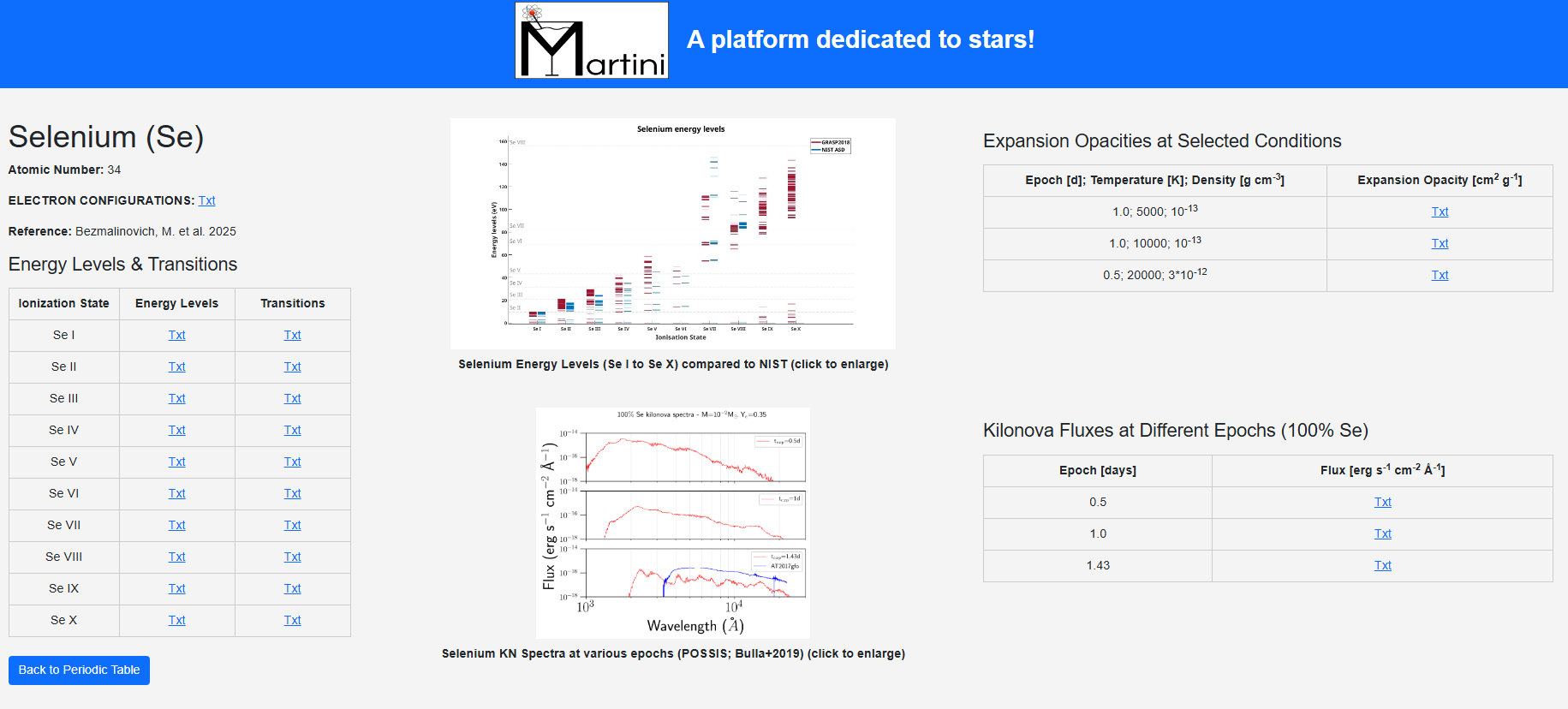}
        \vspace{0.5ex}
        \small \textbf{(c)} Specific data for the selected element.
    \end{minipage}
    \caption{Images of the \texttt{MARTINI} platform. \textbf{(a)} the homepage of the platform with all the different fields of interest, from s-process to KN light curves. \textbf{(b)} the periodic table (showing up) as intermediate window after the topic selection. Here the user must select one of the available element (in green) to move to the data page. \textbf{(c)} a detailed Se example of the data page. In this page there are reported all the literature references, the ``.txt'' data files and plots.}
    \label{fig: MARTINIPlatform}
\end{figure*}

\texttt{MARTINI} is a flexible, user-friendly, open access platform dedicated to astrophysics, in particular to element nucleosynthesis. It combines html pages with a light {\it mysql} database, all embedded in the high-level Python Web framework Django. \reffig{fig: MARTINIPlatform} shows the design and structure of the platform. In detail, all the main topics (e.g., s-process yields, previously available from the \texttt{FRUITY} database website \citep{cristallo2011,cristallo2015} and here now integrated, r-process, AGB dust, atomic opacities, kilonova spectra, and kilonova light curves) are displayed on the homepage (\reffig{fig: MARTINIPlatform}~\hyperlink{fig: MainMenu}{(a)}). After selecting the field of interest for this article (Atomic-Opacities), a new window is loaded, showing the periodic table of the elements (\reffig{fig: MARTINIPlatform}~\hyperlink{fig: PeriodicTable}{(b)}). In the periodic table, green buttons represent the available elements with corresponding atomic data, and red the in-progress elements, while transparent buttons represent the ones not available yet. Once the element is selected, a new page is shown with all the corresponding ``.txt'' data files, i.e.: electron configurations, energy levels and transitions for ground and ionised states, opacities and fluxes at various epochs (\reffig{fig: MARTINIPlatform}~\hyperlink{fig: AtomicDetail}{(c)}). On top of the page, the general properties together with the corresponding literature references associated to the data of the searched element can be found. Plots of selenium energy levels (compared to \texttt{NIST ASD}) and KN spectra at various epochs (computed with \texttt{POSSIS}) are reported as well.\\
The same scheme will be maintained for the future computation of atomic opacities related to other chemical elements.

\section{Conclusions}

To investigate and analyse the impact of selenium in the KN spectra in the early-stage scenario, in this work we performed atomic calculation and opacity estimation for Se I-X ionisation states. The atomic data are calculated using the \texttt{GRASP2018} code, and the results are compared with the \texttt{NIST ASD} and other existing data available from the literature \citep{Tanaka2020,Radit2022,Kitovien2024}. Specifically, Se I-IV energy levels deviate $\sim$ 5\% from the \texttt{NIST ASD} which results in a significant improvement compared to \citet{Tanaka2020}. However, our atomic data do not reach the same accuracy level of \citet{Radit2022} and \citet{Kitovien2024}, suggesting that future refinements in atomic calculation are needed. Concerning Se V-VIII, the precision in the calculations is even higher than the Se I-IV results. In detail, Se V and Se VII are $\sim$ 2,5\%, while Se VI and Se VIII show deviations less than 2\%. Since \texttt{NIST ASD} atomic data are not provided for Se IX-X degrees of ionisation, here the reliability of the results depends only on the quality of the level transitions calculated through a classification method described in \citet{Kitovien2024}. The results show a high transition accuracy for ionisation stages beyond Se IV. Taking into account the high quality of these transitions (\reffig{fig: Full_Transition}), together with the high accuracy achieved on the energy levels (\reftab{tab: ErrorEstimation_Energy_SeV-X}), we provide a robust estimation for Se expansion opacity under the LTE regime. The analysis is performed assuming different temperatures (e.g., $\mathrm{T}=5\,000\,\mathrm{K}$, $\mathrm{T}=10\,000\,\mathrm{K}$, $\mathrm{T}=20\,000\,\mathrm{K}$, and $\mathrm{T}=100\,000\,\mathrm{K}$). The estimated expansion opacity is calculated with improved accuracy compared to existing works in the literature \citep[e.g.,][]{Tanaka2020}. In the last part of the work, we implemented new grids of opacity in the \texttt{POSSIS} code to perform spectral analysis. The grids were calculated through the Se atomic results discussed in \refsec{sec: SeResults} and a new set of densities and temperatures, ranging from -19.5 to -4.5 $\mathrm{g\,cm^{-3}}$ in log-scale and from 1 000 to 51 000 K, respectively (see \refsec{sec: SpectraAnalysis} for more details). In the analysis of the spectra, different KN epochs (e.g., $t_{\mathrm{exp}}=0.5\,\mathrm{d},\,1\,\mathrm{d},\,1.43\,\mathrm{d}$) were investigated, assuming two different scenarios: the first in which the opacity contribution comes from an ejecta made of 100\% selenium, and the second in which the opacity of Se contributes only to 10\% of the total KN mass, while the rest comes from an adopted gray opacity of $0.5\,\mathrm{cm^2\,g^{-1}}$. From the results of the KN spectra, Se spectral features can be observed only in the case of 100\% of the KN composition, while pushing to a more realistic scenario of 10\%, these features are no longer detectable. Finally, all the selenium results discussed in this paper are now available in the new \texttt{MARTINI} platform dedicated to astrophysics, in particular to element nucleosynthesis. These new Se data offer valuable input to the community to refine future KN models and opacity studies, and they should be considered also to re-investigate previous results.

\begin{acknowledgements}
      The authors thank the reviewers for their constructive feedback, which significantly improved the clarity and quality of this manuscript. M. Bezmalinovich, D.V. and S.C. acknowledge funding by the European Union – NextGenerationEU RFF M4C2 1.1 PRIN 2022 project ``2022RJLWHN URKA'' and by INAF 2023 Theory Grant ObFu 1.05.23.06.06. M. Bulla acknowledges the Department of Physics and Earth Science of the University of Ferrara for the financial support through the FIRD 2024 grant. M. Bezmalinovich, D.V. and S.C. are also grateful to S. Simonucci, S. Taioli, T. Morresi, J.H. Gillanders and M. McCann for their support during the work. Finally, the authors acknowledge the support of the \texttt{PANDORA} INFN collaboration.
\end{acknowledgements}

%
%

\bibliography{Bibliography}{} 
\bibliographystyle{aa} 

\begin{appendix}
\onecolumn
    
\section{Additional figures}    
\begin{figure*}[ht!]
    \centering
    \includegraphics[width=0.7\linewidth]{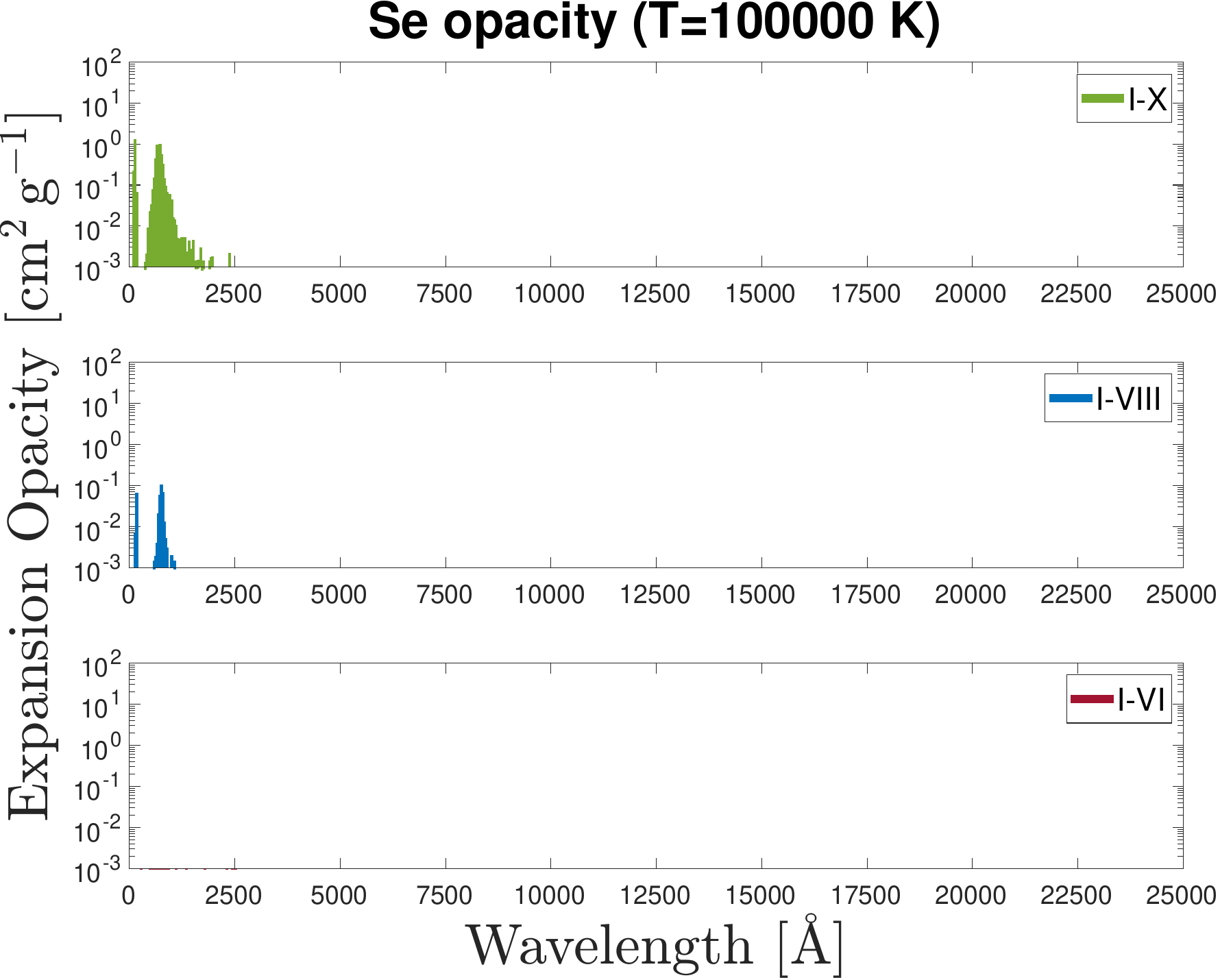}
    \caption{Comparison of Se I-X expansion opacity at $\mathrm{T}=100\,000\,\mathrm{K}$. In the panels, it can be seen that the only opacity contribution at $\mathrm{T}=100\,000\,\mathrm{K}$ comes from the ionisation stages beyond Se VI. No significant contribution can be observed in the ionisation stage between Se I-VI.}
    \label{fig: SeI-IV_full_opacity_T100000}
\end{figure*}

\begin{figure*}[ht!]
    \centering
    \includegraphics[width=0.9\linewidth]{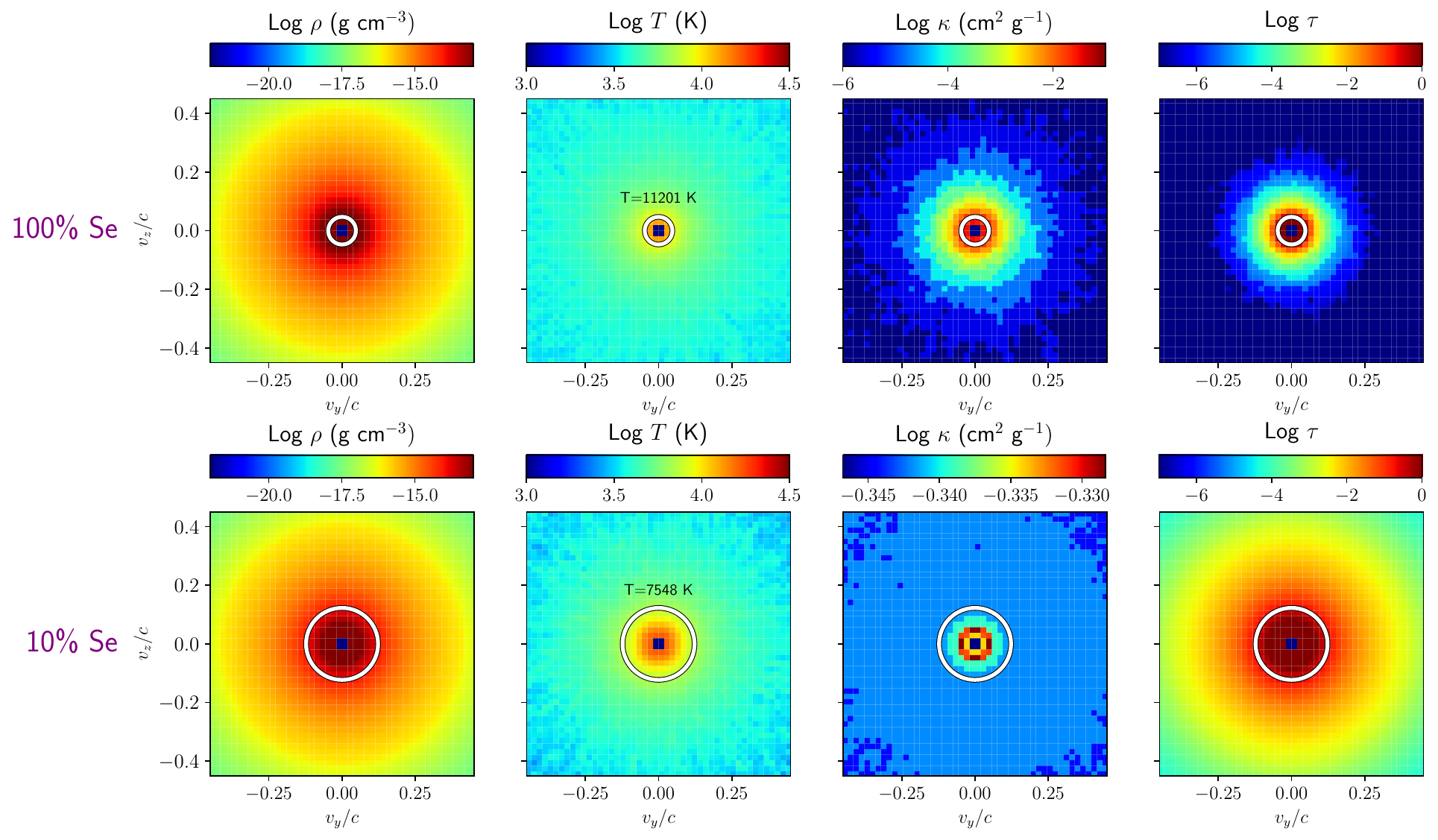}
    \caption{Density, temperature, opacity and optical depth (from left to right) maps in the $yz$ plane for the model with 100\% Se (upper panels) and 10\% Se (lower panels). The white circles show the location of the thermalisation/photospheric surface, and the value of its temperature is indicated in the corresponding panels. Maps are computed at 1\,d after the merger. Note that the colourbars of the upper and lower panels have the same ranges, except for those of the opacities $\kappa$.}
    \label{fig: Tph}
\end{figure*}
\twocolumn
\end{appendix}

\end{document}